\documentclass[aip,jmp,amsmath,amssymb,nofootinbib,reprint]{revtex4-1}
\usepackage{float}
\usepackage{graphicx}%
\usepackage{dcolumn}%
\usepackage{bm}
\usepackage{subfig}
\usepackage{color}
\begin{document}

\title[The culmination of an inverse cascade: mean flow and fluctuations]{The culmination of an inverse cascade: mean flow and fluctuations} 

\author{Anna Frishman}
 
\affiliation{$^1$Princeton Center for Theoretical Science, Princeton University, Princeton, New Jersey 08544, USA}

\date{\today}

\begin{abstract}
Two dimensional turbulence has a remarkable tendency to self-organize into large, coherent structures, forming a mean flow. The purpose of this paper is to elucidate how these structures are sustained, and what determines them and the fluctuations around them. A recent theory for the mean flow will be reviewed. The theory assumes turbulence is excited by a forcing supported on small scales, and uses a linear shear model to relate the turbulent momentum flux to the mean shear rate. 
Extending the theory, it will be shown here that the relation between the momentum flux and mean shear is valid, and the momentum flux is non-zero, for both an isotropic and an anisotropic forcing, independent of the dissipation mechanism at small scales. 
This conclusion requires taking into account 
that the linear shear model is an approximation to the real system.
The proportionality between the momentum flux and the inverse of the shear can then be inferred most simply on dimensional grounds. Moreover, for a homogeneous pumping, the proportionality constant can be determined by symmetry considerations, recovering the result of the original theory. 
The regime of applicability of the theory, its compatibility with observations from simulations, a formula for the momentum flux for an inhomogeneous pumping, and results for the statistics of fluctuations, will also be discussed.  
\end{abstract}

\maketitle
\section{Introduction}
As all physicists know, the two-dimensional world is full of surprises. One such beautiful surprise is the inverse turbulent cascade---the transfer of kinetic energy, in a two dimensional fluid, to scales larger than the initial perturbation\cite{kraichnan_inertial_1967,fjortoft_changes_1953}.  This phenomenon is most striking in a finite domain, where it can lead to the self organization of the flow into a large scale, coherent structure. That the inverse cascade can culminate in a large scale flow was first understood by Kraichnan\cite{kraichnan_inertial_1967}. Relying on the intuition of a Bose-Einstein condensate in equilibrium statistical mechanics, Kraichnan predicted that energy would condense in the largest available Fourier mode. 
Though not strictly correct\cite{frishman_jets_2017}, this has been a lasting and important idea, underlying the basic thinking about the resulting large scale flow.
Qualitatively, one expects the flow to occupy 'the largest available scale', and to be shaped by the geometry of the domain.  

Part of the interest in coherent structures supported by turbulence in (quasi) two dimensions, stems from atmospheric and oceanic dynamics, where such structures are ubiquitous and it is believed that turbulence plays a significant role\cite{vallis_atmospheric_2006}. The emergence of coherent structures in two dimensional trubulence has been verified already in the early laboratory and numerical studies\cite{sommeria_experimental_1986,smith_bose_1993}. 
Moreover, there is evidence that such flows have a universal character\cite{xia_spectrally_2009,chertkov_dynamics_2007,laurie_universal_2014}.

The present work aims to explore and elucidate what determines these coherent flows and the fluctuations around them, in what sense they are universal, as well as precisely in what range of parameters the answers to these questions are given. The text both reviews existing results, presented in a light that will hopefully be illuminating, and introduces some original work. It does not, however, attempt to review all existing literature on mean flows in two dimensional turbulence.

We will consider the dynamical steady state reached by the two dimensional Navier-Stokes equation with linear friction (drag) and forcing. The coherent, or mean, flow is obtained by averaging over time. 
Section \ref{sec:2} introduces the setting, and discusses the relevant non-dimensional parameters and their assumed limits. Section \ref{sec:mean} is devoted to determining the emergent mean velocity. The presence of the strong mean flow is in fact an advantage from a theoretical point of view, since it suppresses non-linear turbulent interactions, making the system much more tractable than turbulence in the absence of such a mean flow. We review the quasi-linear approximation which takes advantage of this property, in section \ref{sec:3}.

Section \ref{sec:vortex} concerns the vortex mean flow, describing the analytic solution for its profile, found recently\cite{laurie_universal_2014,falkovich_interaction_2016}. The derivation relies on a particular closure of the energy balance, which connects
the Reynolds stress (turbulent momentum flux $\langle uv\rangle$) to the mean shear rate, $U'$. For a homogeneous pumping the relation reads $U' \langle uv\rangle =\epsilon$, i.e the left hand side---the local energy exchange between mean flow and fluctuations, is equal to the local kinetic energy injection rate $\epsilon$.
The most significant original result of the present work is section \ref{sec:uvU}, which elucidates the origin of this relation,
building on the works\cite{kolokolov_structure_2016,woillez_first_2016}. It is argued that, in the regime considered here, the momentum flux remains non-zero whether the forcing is isotropic or not. This is independent of the main dissipation mechanism at the forcing scale, and so applies to the regime that seems most relevant\cite{bouchet_statistical_2012}---where bottom drag dominates over viscosity. We further claim that the relation $U' \langle uv\rangle =\epsilon$ holds only for a homogeneous pumping. Indeed, it can be derived from an additional symmetry present for such a forcing. For an inhomogeneous pumping, the Reynolds stress is again proportional to the inverse of the shear rate, but can have a more complicated dependence on the energy injection profile. As an example, the Reynolds stress for a rapidly varying energy profile is treated at the end of section \ref{sec:uvU}. 

Section \ref{sec:validity} briefly discusses the regions of the domain where the universal vortex solution is expected to hold, commenting on the global energy and enstrophy balances. Section \ref{sec:jets} is devoted to another simple setting for the mean flow---that of jets in a periodic domain. The approximation of section \ref{sec:uvU} is applied to this case and the prediction for the mean flow derived.  The results obtained in direct numerical simulations in this setting\cite{frishman_jets_2017} are then described, and an attempt to reconcile the two is made. 

Once the mean velocity profile is established, it becomes possible to obtain the velocity-velocity correltion functions, and hence also the average turbulence energy level. This is the subject of section \ref{sec:fluct}, which focuses on the vortex geometry. It is asserted that, unlike the Reynolds stress, the energy of the fluctuations is determined by the zero modes of an (mean-flow) advectivion equation, also called the Lyapunov equation. The derivation of this equation, and its solutions, are reviewed in this section. A full discussion of the results and comparison to numerical simulations are deferred to an upcoming work\cite{herbert_fluctuations_2017}. 

Finally, a summary and discussion of the emergent picture for the mean flow and fluctuations, which are the culmination of the inverse cascade, is presented in \ref{sec:smry}. 

\section{Parameters of the problem}
\label{sec:2}
The 2D Navier-Stokes equation with linear drag and forcing reads,
\begin{equation}
\partial_t \bm{V} +\bm{V}\cdot \nabla \bm{V}=-\nabla P -\alpha \bm{V} +\bm{f} -\nu (-\Delta)^p \bm{V} 
\label{eq:NS} 
\end{equation}
where we use hyper-viscosity, also frequently used in numerical experiments, with the usual viscosity recovered for $p=1$. 
We decompose the velocity $\bm{V}=\bar{V}+\bm{v}$ into a mean, $\bar{V}$ and fluctuating part $\bm{v}$, where the average is taken over time, or equivalently (assuming the dynamics to be ergodic), over realizations of the forcing at statistically steady state. For the fluctuations, averaging will be denoted by angular brackets $\langle\cdot \rangle$. In the following, we will assume throughout that the forcing has a zero average and is white in time correlated. Some comments on a finite-time correlated forcing and on time averaging in a periodic box can be found in Appendix \ref{app:cor_time}.

In addition to the kinetic energy injection rate, $\epsilon$, the forcing wavenumber $q_f$ and the viscosity $\nu$---parameters characterizing three dimensional turbulence---the two dimensional system contains also the box size $L\gg 2\pi/q_f\equiv l_f$, for an inverse cascade setting, and the linear friction rate $\alpha$. The latter is used to model large scale dissipation, which is always present in experiments (bottom drag for example), and insures the system reaches a steady state in a feasible time for numerical simulations.
There are therefore three non-dimensional parameters to be used to characterize a given system. The first is $\delta \equiv \alpha (L/2\pi)^{2/3}\epsilon^{-1/3}$, corresponding to the ratio of the inverse cascade timescale to that of friction.  Next is the ratio $K\equiv L q_f/2\pi$ between the forcing scale and the size of the box. Finally, 
defining the rate of viscous dissipation at the forcing scale $\gamma=\nu q_f^{2p}$, we have $\Gamma = \gamma/\alpha$ which determines the relative strength of viscous dissipation and friction for the direct cascade (i.e the scales below the forcing scale). 

To initiate an inverse cascade, dissipation at the forcing scale must be inefficient compared with the non-linear transfer, implying  $\Gamma ^{-1} \delta^{-1} K^{2/3} \gg 1$ and $\delta^{-1} K^{2/3}\gg 1$. Furthermore, in order for a mean flow to form, the energy extraction at the domain scale should be much slower than the non-linear transfer, i.e $\delta \ll 1$. Here and in the following we are assuming that at the system scale friction is faster than viscosity:  $\Gamma K^{-2p} \ll 1$. It appears that both simulations and experiments are in this regime.
Finally, we expect that as the forcing correlation length is decreased the emerging mean flow would become less dependent on the forcing features, and hence we will work in the asymptotic limit $K\to \infty$. 
These relations are summarized in Table \ref{table:1}.
Of course, as we are concerned with the statistical steady state, the limit $t \to \infty$ is assumed to be taken first.
 
\begin{table}
\centering
\begin{tabular}
{| c | c | c | c | c | c | c |}
\hline   & & & & & &\\
$\Gamma$ & $\delta\ll1$ & $K\gg1 $ &  $\frac{\Gamma}{ K^{2p}}\ll1$  &  $\frac{\Gamma\delta}{ K^{2/3}}\ll1$ & $d/R_u\ll 1$  & $\Lambda\gg 1$\\ & & & & & &\\\hline & & & & & &\\
  	 $\frac{\nu q_f^{2p}}{\alpha}$ & $\frac{\alpha L^{2/3}}{2\pi\epsilon^{1/3}}$ & $\frac{L q_f}{2\pi}$ &  $\frac{\nu}{ L^{2p}\alpha}$ & $\frac{\nu q_f^{2p-2/3}}{\epsilon^{1/3}}$ & $ \frac{d \sqrt{\delta} K^{2/3}}L $ & $ \frac{q_f d}{2\pi} $\\ & & & & & &\\	\hline 
\end{tabular}
\caption{Dimensionless combinations of the parameters of the problem. Throughout the text we assume the limits stated in the first line are satisfied. The parameter $d$ denotes a characteristic length scale for the mean flow gradient.}
\label{table:1}
\end{table}

It is useful to estimate the magnitude of the mean velocity from the energy balance. Since energy goes to large scales, we expect that in the steady state the energy injected by the forcing is balanced to leading order by the energy dissipated by the mean flow.  
As we have assumed the balance is dominated by friction, it gives the estimate $|\bar{\bm{V}}| \propto \sqrt{\epsilon/\alpha}$. We can also define a typical time associated with the mean flow, using the box scale $L$: $\tau_m =\sqrt{\alpha/(L^{-2} \epsilon)}$.

\section{Predicting the mean flow}
\label{sec:mean}
The most basic question one could ask is---what determines the large scale coherent pattern into which the flow organizes itself?
Qualitatively, Kraichnan's idea of condensation of energy implies that the flow would form the largest coherent structure that conforms with symmetries (of the domain). In addition, the scale the mean flow eventually occupies, should be able to support a global balance between dissipation and injection of momentum and vorticity.
However, these considerations alone do not give a detailed prediction for the resulting mean flow, and even this qualitative picture does not seem to always work \cite{frishman_jets_2017}.

Another approach, is to use equilibrium statistical mechaincs of two dimensional fluid mechanics \cite{miller_statistical_1990,robert_statistical_1991,bouchet_statistical_2011}.
It remains unclear how to use this approach to predict the emergent mean flow for a forced flow (where the distribution of vorticity in the steady state is not known  a priori), nor whether it is justified in this case to apply an equilibrium theory to an out-of-equilibrium system. 

Instead, in the following we will focus on an approach that utilizes the dominance of the mean flow over the fluctuations, called the quasi-linear approximation. 
\subsection{The quasi-linear approximation}
\label{sec:3}
The quasi-linear approximation is a way to close the hierarchy of equations for the velocity (or vorticity) moments. It relies on the presence of a strong mean flow, which implies that non-linear interactions between fluctuations may be neglected. This approach has been widely used to study the statistical steady state\cite{nazarenko_exact_2000,farrell_structure_2007,marston_statistics_2008,bouchet_kinetic_2013}  as well as a reduced model for the systems dynamics (particular cases being the so called S3T or CE2) \cite{farrell_structural_2003,srinivasan_zonostrophic_2011,parker_generation_2014}, to give a non-exhaustive list of related literature (this approach is also similar to RDT\cite{hunt_rapid_1990}). The dynamical approach has been particularly popular in atmospheric dynamics, where one is interested in the formation and merging of large scale jets in the presence of differential rotation. 
This will not be the focus of the present work, but let us mention that using CE2 or S3T requires a time separation between the (slow) evolution  of the mean flow and the (fast) evolution of fluctuations \cite{bouchet_kinetic_2013,bouchet_stochastic_2014}. 

While the dynamical approach is tractable numerically, the quasi-linear approximation remains too complicated to allow an analytic solution for the steady state mean flow. 
We will now describe the quasi-linear approach in some detail, and in the next section present an additional step, first used in \cite{laurie_universal_2014},  which makes an analytic solution for the mean flow possible.

The basic idea is to start with the Reynolds averaged Navier-Stokes equation. As usual, it relates the mean velocity to the Reynolds stress---a single point velocity-velocity correlation function. The next equation in the hierarchy is for the velocity covariance, with velocities taken at two different spatial points. It would depend both on the mean velocity and on higher order correlation functions. A closure is then adopted at second order, throwing away third order and higher order terms. Qualitatively, these non-linear interactions between fluctuations do not have time to influence the dynamics if the mean flow shear acts much faster. So, we require that the mean shear rate at a given point, $U'\sim U/d$ where $d$ is a characteristic length scale, is much larger than the non-linear transfer rate at the forcing scale $\sim \epsilon^{1/3}q_f^{2/3}$. This gives the condition $d/L \ll \delta^{-1/2}K^{-2/3}\equiv R_u/L$, 
in addition to the other limits listed in Table \ref{table:1}. Note that generically we also expect $\Lambda \equiv q_f d/2\pi \gg 1$, which will be important later on. 
If $R_u<L$ then $R_u$ will dictate the range of radii where the quasi-linear approximation is applicable\cite{kolokolov_structure_2016}. 
Importantly, we see that once the forcing scale is assumed small, the approximation is no longer simply controlled by $\delta\to 0$.

Let us now present the formal derivation of the equations in the quasi-linear approximation. We begin by averaging over equation (\ref{eq:NS}), using the assumptions of Table \ref{table:1}. In addition to neglecting the dissipation terms, we may assume $\nabla \cdot \langle \bm{v} v^j \rangle \ll \bar{\bm{V}}\cdot \nabla  \bar{V}^j$, since most of the energy is contained in the mean flow. Then, at leading order, we get that the mean velocity is a solution of the steady Euler equation $\bar{\bm{V}}\cdot\nabla  \bar{\bm{V}} =-\nabla \bar{P}$, or $\bar{\bm{V}}\cdot\nabla  \bar{\Omega} =0$ for the mean vorticity $\bar{\Omega}$. 
Usually, one considers a setting where this equation is satisfied to all orders due to the symmetries of the mean flow: if the mean velocity depends only on the coordinate transverse to its direction, the equation is satisfied automatically. In that case, the sub-leading terms of the mean velocity need not to be considered separately for the next order.

To be concrete, let us focus on a polar mean flow which depends only on the radius: $\bar{\bm{V}}=(U(r),0)$, relevant for a vortex\cite{laurie_universal_2014}, and denote the polar and radial velocity fluctuations by $\bm{v}=(u,v)$. We assume the system is statistically isotropic, i.e independent of the polar angle $\phi$, see Figure \ref{fig:cond}~c). The ensuing discussion straightforwardly translates to a different geometry, e.g a jet with $\bar{\bm{V}}=(U(y),0)$.

The averaged (angular) momentum balance at the next order (since $\Gamma K^{-2p} \ll 1$) gives  
\begin{equation}
r^{-1}\partial_r (r^2 \langle uv \rangle) =-\alpha r U.
\label{eq:momentum}
\end{equation}

For the fluctuations, instead of writing the equation for the velocity covariance, it would be more convenient to work with a dynamical equation and compute the steady state correlation functions from it.
The dynamics of the fluctuations in the background of the mean flow, written for the vorticity $\omega =\nabla \times \bm v$, reads 
\begin{equation}
\partial_t \omega+(U+u)/r \partial_{\phi} \omega  +v \partial_r (\Omega + \omega) =g -\alpha \omega -\nu (-\Delta)^{p} \omega
\label{eq:fluct_vorticity}
\end{equation}
where $\Omega = r^{-1} \partial_r (r U)$ and $g=\nabla\times \bm{f}$.
Neglecting terms non-linear in fluctuations we have:
\begin{equation}
\partial_t \omega+U/r \partial_{\phi} \omega  +v \partial_r \Omega =g -\alpha \omega  -\nu (-\Delta)^{p} \omega.
\label{eq:fluct_vorticity_app}
\end{equation}

In principle, equations (\ref{eq:momentum}) and (\ref{eq:fluct_vorticity_app}) form a closed system: we can express $u,v,\omega$ in terms of the stream function. The Reynolds stress $\langle uv\rangle$ can then be expressed as an integral over the stream function covariance. Finally, the steady state equation for the stream function covariance can be written using (\ref{eq:fluct_vorticity_app}), substituting the mean velocity by the Reynolds stress with the help of (\ref{eq:momentum}).   
The resulting equation would be too complicated to produce an expression for $U$, as one would have to solve a non-linear and non local equation for the stream function covariance. Instead, in S3T or CE2 the  dynamical equations for the two point correlation function together with a dynamical version of (\ref{eq:momentum}) are solved numerically.

\textit{Applicability of the approximation}\\
In theory, in the regime of Table \ref{table:1}, both the dissipation and non-linear terms should be negligible in equation (\ref{eq:fluct_vorticity}), although we have kept the former in (\ref{eq:fluct_vorticity_app}). Dissipation of fluctuations is necessary to sustain a steady state for the vorticity: as the large scale mean flow can contain only a small portion of the injected enstrophy (squared vorticity), injection must be balanced by the dissipation of fluctuations, even in the limit of small dissipation and strong mean flow\footnote{If $\Gamma\gg1$ then a dissipative anomaly would be present: \unexpanded{$\nu \langle \nabla \omega_1\nabla \omega_2\rangle \xrightarrow[ \nu\to 0,1\to2]{} \epsilon q_f^2 $}, and for $\Gamma\ll1$ we expect \unexpanded{$\alpha \langle \omega_1 \omega_2\rangle \xrightarrow[ 1\to2]{} \epsilon q_f^2 $} even in the limit $\alpha \to 0$.} (see also the dicsussion in\cite{bouchet_kinetic_2013}). 
On the other hand, we expect most of the energy to be dissipated by the mean flow, so that dissipation of fluctuations need not play an important role for velocity correlation functions\footnote{If the advection by the mean flow is identically zero, like for the velocity zeroth harmonic as discussed in section \ref{sec:fluct}, we may expect non-linear terms to dominate over dissipative ones.}. 

The interested reader may find the comparison of non-linear terms to advection terms, and a discussion of when the former is negligible, in Appendix \ref{app:non-linear}. Here, we only comment that what complicates the evaluation of the range of validity of the quasi-linear approximation, is that it is currently unclear how the (velocity) fluctuations scale with the dimensionless parameters.
Indeed, while relation (\ref{eq:momentum}) produces the estimate $\langle uv \rangle/U^2\propto \delta^{3/2}$ (using $L$ as a characteristic length scale for the mean velocity), one should be careful using this estimate for the fluctuations, as done in
\cite{bouchet_kinetic_2013}. Symmetry considerations imply that the Reynolds stress $\langle uv \rangle$ can be much smaller than, for example, the energy $\langle u^2+v^2 \rangle$\cite{laurie_universal_2014,kolokolov_structure_2016}.
Indeed, without forcing and dissipation, the mean flow and equation (\ref{eq:NS}) are both invariant under the parity+time reversal (PT) transformation $\phi \to -\phi$ and $t\to -t$  while $\langle uv \rangle$ changes sign\footnote{contrary to\cite{kolokolov_structure_2016} we assert that the forcing breaks $t\to -t$, since, for example, in its presence and without dissipation enstrophy  cannot reach a steady state}. 
A forcing with a small-scale correlation length is irrelevant for most scales, so we may expect it to give a sub-dominant contribution to velocity correlation functions.  
Therefore,   
in the limit of Table \ref{table:1}, such that energy dissipation by friction and viscosity for the fluctuations is also negligible, the PT symmetry for the fluctuations could plausibly be restored, and $\langle uv \rangle$ suppressed with respect to $\langle u^2+v^2 \rangle $.
\subsection{Recent progress - the vortex profile}
\label{sec:vortex}

\begin{figure}
\includegraphics[width=\linewidth]{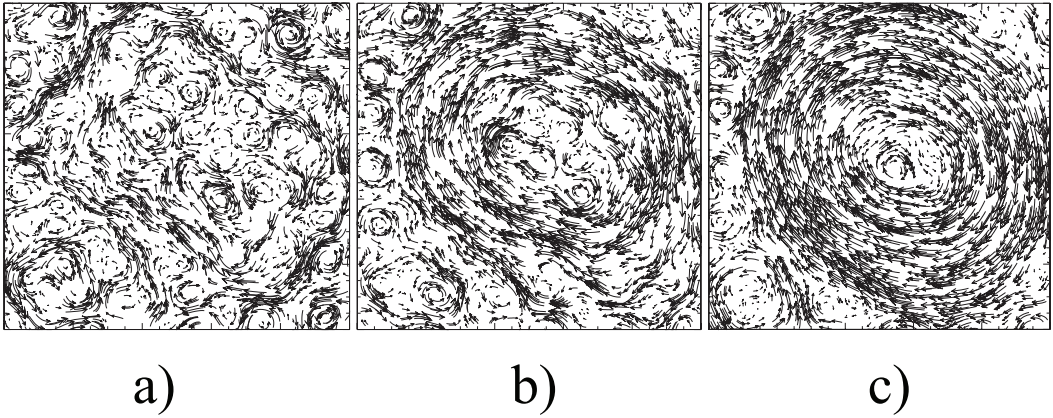}
\caption{Time-averaged velocity fields from a laboratory experiment of turbulence in a thin fluid layer. The container is a square box of side length L=0.1m, the measurements are taken at different damping rates $\alpha$: (a)  $\alpha=0.25s^{-1}$, (b) $\alpha=0.15s^{-1}$, and (c) $\alpha=0.05s^{-1}$.
Reprinted from\cite{xia_spectrally_2009}, with the permission of AIP Publishing.}
\label{fig:cond}
\end{figure}

It has been recently discovered\cite{laurie_universal_2014}, that the momentum flux $\langle uv \rangle$ can be directly and simply related to the mean velocity, simplifying the system (\ref{eq:fluct_vorticity_app}),(\ref{eq:momentum}). Then, with the use of equation (\ref{eq:momentum}) the mean velocity profile can be obtained.   
The setup used was that of a single vortex, which should be applicable both to the large vortex observed in experiments\cite{xia_spectrally_2009}, see Figure \ref{fig:cond}, and to either one of the vortices of the dipole appearing in a periodic square domain\cite{chertkov_dynamics_2007}. 

The starting point is the energy balance for the fluctuations, obtained by subtracting the energy balance of the mean flow from the total energy balance (it can equivalently be obtained from (\ref{eq:fluct_vorticity})) 
\begin{equation} 
\begin{split}
& {1\over r}{\partial\over\partial r}r\left\langle\! v\!\left(\!{u^2+v^2\over2}+p\right)\!\right\rangle =   \\  &= \epsilon-\langle uv\rangle r\partial_r{U\over r}-\alpha\langle u^2+v^2\rangle+ \nu (-1)^{p+1} \langle\bm{v}\cdot\nabla^{2p}\bm{v}\rangle .
\end{split}
\label{eq:energy}
\end{equation}
The left hand side is the spatial energy flux, which is balanced on the right hand side by the energy injection rate, the dissipation due to friction and viscosity and $\langle uv\rangle r\partial_r{U\over r}$, the product of the angular momentum flux and the angular velocity shear rate. This is the energy exchange between the mean flow and the fluctuations.

Next, applying the quasi-linear approximation, the dissipation and the cubic-in-velocity terms can be neglected. This still leaves the pressure on the left hand side of Eq. (\ref{eq:energy}), since it contains contributions which are a product between the mean and fluctuating velocity. The next bold step, going beyond the quasi-linear theory, is to throw away this term and obtain:
\begin{equation} 
\langle v u\rangle r= \frac{\epsilon}{\partial_r{U\over r}}  
\label{eq:uv}
\end{equation}
The energy exchange between the mean flow and fluctuations thus serves as a local mechanism of energy removal from the fluctuations. In addition, since $\langle uv\rangle r\partial_r{U\over r}=\epsilon>0$ angular momentum flows to regions with large mean angular velocity, which is the opposite of a diffusive situation (bringing to mind the idea of a negative eddy viscosity\cite{kraichnan_eddy_1976} ).

Now, the system (\ref{eq:momentum}), (\ref{eq:uv}) is solved by substituting a power law solution for the mean velocity, giving
 \begin{equation}
 U=\sqrt{3\epsilon/\alpha}\quad \langle uv\rangle=-r\sqrt{\epsilon\alpha/3} \qquad \langle p\rangle = 3\epsilon/\alpha \ln(r/R)\ 
 \label{eq:vortex_sol}
 \end{equation}
a universal result, seemingly independent of the forcing details and boundary conditions. 
 
\begin{figure*}
\subfloat[]{\includegraphics[width=0.3\linewidth]{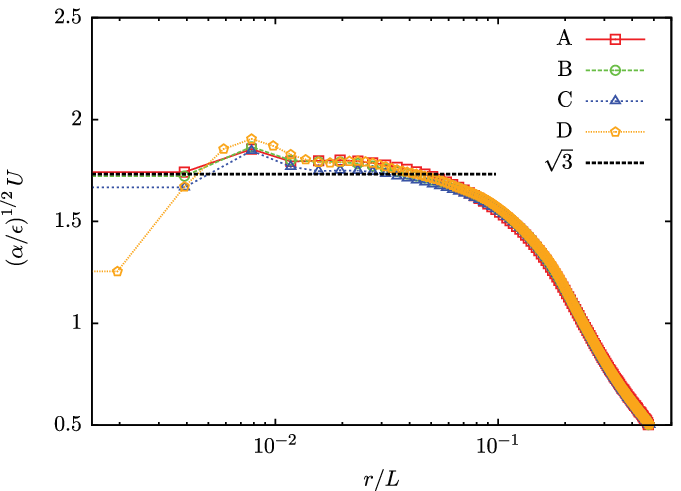}}
\subfloat[]{\includegraphics[width=0.3\linewidth]{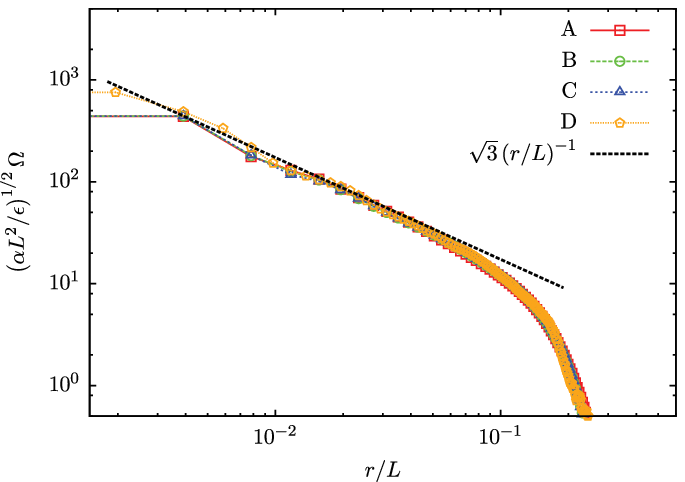}}
\subfloat[]{\includegraphics[width=0.28\linewidth]{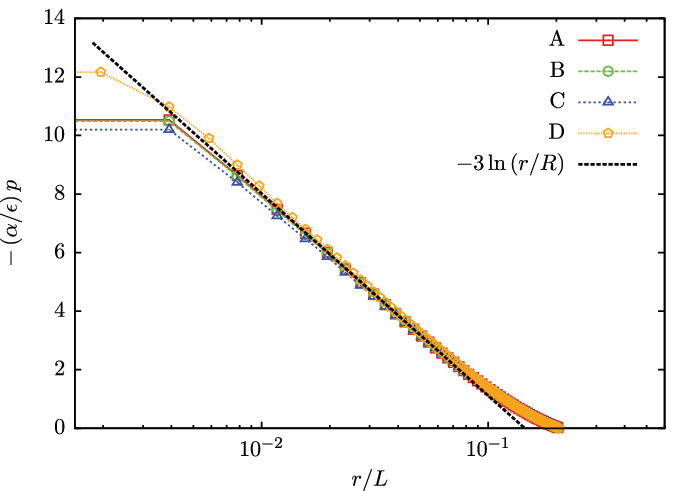}}
\caption{a) Radial profile of the mean velocity. 
b) Radial profile of the mean vorticity $\Omega$. c) Radial profile of the mean pressure. In all three figures the black dashed line corresponds to the theoretical prediction (\ref{eq:vortex_sol}).
Reprinted figure with permission from\cite{laurie_universal_2014} \copyright (2014) by the American Physical Society.}
\label{fig:num}
\end{figure*}

A remarkable agreement was found between this prediction for the mean flow and pressure, and results from numerical simulations\cite{laurie_universal_2014}, see Figure \ref{fig:num}. The mean velocity profile, equation (\ref{eq:vortex_sol}) implies a mean vorticity of the form $\Omega\propto 1/r$. This agrees well with simulations, Figure \ref{fig:num} b), and roughly agrees with preliminary results from experiments\cite{xia_spectrally_2009}.  For the pressure, the value $R/L=0.143$ was extracted from the numerical data, Figure \ref{fig:num} c), giving an estimate of the vortex size. Validation of the solution (\ref{eq:vortex_sol}) for the Reynolds stress is more challenging\cite{herbert_fluctuations_2017}.

\subsection{Connecting the Reynolds stress to the mean velocity}
\label{sec:uvU}
The relation (\ref{eq:uv}) is simple and beautiful, but one has to wonder when and why it is satisfied. Building on the ideas of Kolokolov and Lebedev \cite{kolokolov_structure_2016},
as well as Woillez and Bouchet \cite{woillez_first_2016}, the aim of the current section is to answer these questions.

The main idea\cite{kolokolov_structure_2016} for the calculation of $\langle uv\rangle$ is to approximate the mean flow locally by a linear shear. Then, a closed form for the momentum flux can be obtained through a balance between the forcing and the shear. Deviating from\cite{kolokolov_structure_2016}, we will argue that the result is independent of the dissipation mechanism, which can be set to zero in the calculation. In particular, $\langle uv\rangle$ is non-zero for an isotropic forcing even at $\Gamma\ll1$. This result seems to contradict\cite{srinivasan_reynolds_2014} where a linear shear with zero viscosity and vanishingly small friction was considered, and $\langle uv\rangle=0$ was found. As we will show below, it is important that the linear shear is only an approximation to the real flow, and this results in a delicate order of limits. It turns out that\cite{srinivasan_reynolds_2014} corresponds to the opposite order of limits than the one we are concerned with here. 

Let us first present a simple line of reasoning that gives relation (\ref{eq:uv}), before we discuss a detailed calculation in support of it. We start with the realization that in the absence of forcing and dissipation, both $\langle uv \rangle$ and $\langle vp\rangle$ would be zero: both change sign under the PT symmetry $t\to -t, \phi\to -\phi$ which the system is invariant under (see the discussion at the end of sec. \ref{sec:3}). We claim that the forcing is the leading order symmetry breaking term that need be considered to compute $\langle uv \rangle$ (and $\langle vp\rangle$). The momentum flux is therefore the result of a balance between advection by the mean flow and the injection by the forcing in equation (\ref{eq:fluct_vorticity}).  
Consequently $\langle uv\rangle$ is dominated by modes at the forcing scale, so that, for most of the relevant modes, the only important feature of the mean flow is the shear rate. 
More generally, it would be only the local dynamics, and locally determined quantities, which would enter for such modes. Thus, we can work in local coordinates---setting the origin at a given radius and angle, and using Cartesian coordinates. We then have the mean flow $U(y)$ pointing in the $x$ direction---which translates to the angular velocity $U(r)/r$ in the $\phi$ direction, and the local shear rate $\partial_y U(y)\equiv U'$ and the Reynolds stress $\langle uv\rangle$---corresponding to the angular shear rate $\partial_r (U/r)$ and the angular momentum $\langle uv \rangle r$. 

Now we can obtain relation (\ref{eq:uv}) by dimensional analysis and symmetry considerations. The dynamics in 
the quasi-linear approximation is linear in the fluctuations and the forcing, and since the solution is determined by the latter, it too is linear in the forcing---implying that the Reynolds stress is proportional to $\epsilon$. Moreover, we are considering the inviscid limit, meaning that the only remaining time scale is $(U')^{-1}$ which gives $\langle uv \rangle \propto \epsilon/ U'$ on dimensional grounds (correspondingly, the time scale $r\partial_r (U/r)$ and the relation $\langle uv \rangle \propto \epsilon/ (r\partial_r (U/r))$ in polar coordinates).

We claim that relation (\ref{eq:uv}), where the proportionality constant is exactly unity, holds only if the forcing is statistically homogeneous in both directions---independent of $y$ ($r$) in addition to $x$ ($\phi$). In this case, the local dynamics as well as the local shear rate $U'$, have the symmetry $x\to -x, y\to -y$ ($\phi\to -\phi$ and reflection with respect to a given radius).  
As $\langle vp\rangle$ is determined by the local dynamics to leading order, and changes sign under this symmetry,
it is zero in the main order (we show this by an explicit calculation in Appendix \ref{sec:vp}). Thus, the energy balance (\ref{eq:energy}) (without non-linear terms and dissipation) gives (\ref{eq:uv}). Note that, unlike $\langle vp\rangle$, $\langle uv \rangle$ is invariant under the $x\to -x, y\to -y$ symmetry and so can remain finite. Also, adding a finite (but small) viscosity to the local dynamics does not alter this picture, since it doesn't break the latter symmetry.\footnote{Adding a finite uniform friction also does not break the symmetry. However, depending on the order of limits taken, the dissipation of fluctuations , rather than \unexpanded{$\langle vp\rangle$}, can become non negligible.} 

To arrive at the above picture one had to first assert that the forcing gives the main symmetry breaking contribution to $\langle uv\rangle$. 
Indeed, it breaks time reversal symmetry: it pumps enstrophy into the system, which would increase with time in the absence of dissipation. Thus, generically it is the first symmetry breaking effect that can become important. Given that in the absence of forcing the Reynolds stress is zero by symmetry, it must be the balance between advection and forcing which determines $\langle uv\rangle$. 

The second important step was to realize that the only quantity of interest is the shear rate. This ingenious insight is at the core of the work\cite{kolokolov_structure_2016}. Let us show how this result comes about through a direct derivation of $\langle uv\rangle$. The starting point is the dynamical equation for the fluctuating vorticity, in the regime of table \ref{table:1} (see discussion at the end of Sec. \ref{sec:3}), written in local coordinates \footnote{For the vortex mean flow, the deviations from a globally Cartesian system are expected to give rise to corrections suppressed by a factor $(rq_f)^{-1}$ which is assumed small in the following}:
\begin{equation}
\partial_t \omega+U(y) \partial_{x} \omega  -v U''(y) = g 
\label{eq:fluct_vorticity_app_c}
\end{equation}

The system is homogeneous in $x$ so that we can work in Fourier space, denoting by $k$ the wavenumber in that direction. We will consider the response, denoted by $\omega_k^l(y)$, to a given Fourier mode of the forcing $(k,l)$: $g_{k}^l(y)=g_{kl} e^{ily}$, where 
\begin{equation*}
\begin{split}
\langle g_{k}^l(y,t)g_{k'}^{l'}(y',t')\rangle  =2\eta_{k,l} e^{il(y-y')}\\
\times (2\pi)^2 \delta(l+l')\delta(k+k')\delta(t-t').
\end{split}
\end{equation*}
The energy injection rate in Fourier space is related to the enstrophy injection rate by $\epsilon_{k,l}=\eta_{k,l}/q^2$ where $q^2=k^2+l^2$.

Notice that $|\partial_t\omega+U(y) \partial_{x} \omega | \gg |v U''(y)| $, which we call the shear approximation, is valid if $U'(y) \gg U''(y) q_f^{-1}$. To arrive at this estimate, we have used that the typical timescale related to advection is $1/U'$ together with the rough estimate $v\sim \omega/q_f $, see a more precise discussion in Appendix \ref{sec:app_shear}. This gives the condition $\Lambda=q_fd/2\pi\gg1$ (recall that $d$ is a characteristic scale for $U$). In what follows, we will assume the limit $\Lambda \to \infty$, so that this approximation applies. For the vortex this restricts the radii to $rq_f \gg1$, the radius $r$ serving as a characteristic scale for $U$, while for a jet we expect $d\approx L$ and $\Lambda \approx K\gg1$.

We therefore obtain the equation 
\begin{equation}
\partial_t \omega_k^l+ iU(y) k \omega_k^l = g_k^l 
\label{eq:w_fluct_min}
\end{equation}
which can be solved to obtain $\langle uv\rangle$.
We will provide a sketch of the calculation for a homogeneous pumping here, and give the full details for an inhomogeneous pumping in Appendix \ref{sec:appB}. 

Following\cite{woillez_first_2016}, 
the velocity is obtained from the vorticity, via the stream function 
\begin{equation*}
(\partial_y^2 -k^2) \psi_{k}(y)=\omega_{k}(y).
\end{equation*}
For a given mode $k\neq 0$, the stream function can be exactly expressed in terms of the vorticity using the Greens function of the equation (see for example\cite{morse_methods_1946}):
\begin{equation}
\psi_{k}^l(y,t)= -\frac{1}{2k^2}\int_{-\infty}^{\infty}dY e^{-|Y|}\omega_{k}^l\left(y-\frac{Y}k,t\right).
\end{equation}

The solution to (\ref{eq:w_fluct_min}) is immediate:
\begin{equation}
\omega_{k}^l(y,t)=\int_{0}^t e^{ik U(y)(t'-t)}g_{kl}(y,t')e^{ily}dt'
\end{equation} 
assuming that the forcing was turned on at time zero, when there was no vorticity in the system $\omega_{k}^l(y,0)=0$.
Our focus is the steady state limit $t\to \infty$. The enstrophy diverges in this limit---in the absence of dissipation it cannot achieve a steady state. However, we are only interested in the Reynolds stress $\langle uv\rangle$, and we will see that it is well defined even in the absence of dissipation. 
Using that 
\begin{align*}
u=-\partial_y \psi && v=\partial_x \psi,
\end{align*}
 averaging over the forcing and changing variables to $\tau=t'-t$ we get, after some manipulations,
\begin{widetext}
\begin{equation}
\langle v_k^l(y) u_{-k}^{-l}(y)\rangle =\frac{i\eta_{k,l} }{2k^2}\int_{-t}^{0} d\tau\int dY e^{-|Y|}\int dY' \left[\partial_{Y'}e^{-|Y'|}\right]   e^{-i\frac{l}k (Y+Y')} e^{-ik U(y+\frac{Y'}{k})\tau}e^{ik U(y-\frac{Y}k)\tau}. 
\label{eq:uv_kl}
\end{equation} 
\end{widetext}
Now comes a key point. So far, we have kept the general dependence on $y$ of the mean flow $U(y)$. We now see that for modes which have $k\gg2\pi/d$ we can expand $U(y-\frac{Y}{k})\approx U(y)-\frac{Y}k U'(y)$ (if $U'\neq 0$) inside the integral in (\ref{eq:uv_kl}), since $Y,Y'\gg1 $ are exponentially suppressed in the integrals. This means that the only aspect of the mean flow which enters the dynamics of modes with $k\gg2\pi/d$ is the shear rate. This discussion demonstrates how the assumption of small scale forcing leads to a localization of the dynamics in the $y$ direction. On the other hand, it shows that for modes with smaller $k$ a local approximation using $U'$ is not possible. We therefore should exclude them from the computation that assumes a linear shear. Recall that $(k,l)$ are the wavenumbers of a particular forcing mode, so that $k^2+l^2 \approx q_f^2$, and this exclusion is approximately equivalent to a restriction on the possible angles 
\begin{equation}
-\Lambda \leq  \tan\theta \leq \Lambda,
\label{eq:lmbda}
\end{equation}
with $\Lambda= q_f d/2\pi$ and $l/k=\tan\theta$. In the limit $\Lambda \to \infty$ almost all modes satisfy the former condition. The contribution to the momentum flux of modes that do not satisfy it, and thus cannot be computed using the linear shear approximation, is therefore expected to be negligible. The existence of the cutoff $\Lambda$, however, will turn out to be important. We will provide a geometric picture for its role below.

So, in the linear shear approximation the contribution from forcing modes with $(k,l)$ is given by
\begin{equation}
\begin{split}
\langle v_k^l(y) u_{-k}^{-l}(y)\rangle & =-\frac{2\epsilon_{k,l}q^2}{k^2}\int_{-t}^{0} d\tau \frac{\frac{l}{k}+U'\tau}{(1+(\frac{l}{k}+U'\tau)^2)^2} .
\label{eq:Leb_uv} 
\end{split}
\end{equation} 

The integrand in (\ref{eq:Leb_uv}) is an exact differential giving, in the limit $t\to \infty$,
\begin{equation}
\langle v_k^l(y) u_{-k}^{-l}(y)\rangle =\frac{\epsilon_{k,l}q^2}{U'k^2} \frac{1}{(1+(\frac{l}{k})^2)}=\frac{\epsilon_{k,l}}{U'} .
\end{equation}

The momentum flux $\langle u v\rangle$ is recovered by integrating over the contributions from all forcing modes $(k,l)$ that satisfy (\ref{eq:lmbda}). Taking $\Lambda \to \infty$ we get $
\langle u(y)v(y)\rangle =\epsilon/U'(y)$ as promised. 

As expected, expression (\ref{eq:Leb_uv}) is also what one obtains in a linear shear mean flow\cite{kolokolov_structure_2016,srinivasan_reynolds_2014}. However, there is a delicate order of limits here that has been overlooked previously. Working in the zero dissipation limit, we first compute the contribution resulting from a given forcing mode which is characterized by $(k,l)$. We integrate over the different excitation times of this mode in the past, and only then preform an integral over all modes. The integral over time produces the response to a given forcing excitation,  which is a measurable quantity in the steady state. Thus, this seems to be the physically sensible order of integration.
Reversing the order of the two integrations with $\Lambda\to \infty$, $t\to\infty$ simultaneously alters the result, as it secretly involves changing the order of limits $\Lambda \to \infty$ and $t\to \infty$ which do not commute. In particular, to get $
\langle u(y)v(y)\rangle =\epsilon/U'(y)$ it is required to take times longer than $|U'|^{-1}\Lambda$. Let us show how this works out for an isotropic forcing where $\epsilon_{k,l}= \chi(q)$. We have (see also the discussion in Appendix \ref{sec:app_noncom})
\begin{equation}
\begin{split}
\langle u v\rangle =-\int \frac{qdq d\theta}{(2\pi)^2}\int_{-t}^{0} d\tau\frac{2\chi(q) }{\cos^2\theta}  \frac{\tan\theta+U'\tau}{(1+(\tan\theta+U'\tau)^2)^2}\\
=-\int \frac{q\chi(q)dq}{\pi}\int_{-\Lambda}^{\Lambda} \frac{dz}\pi \int_{-t}^{0} d\tau \frac{z +U'\tau}{(1+(z +U'\tau)^2)^2}= \\
=\frac{\epsilon}{U'\pi}(2\arctan \Lambda -\arctan (\Lambda-U't)-\arctan (\Lambda+U't))
\end{split}
\label{eq:uv_iso}
\end{equation}
Taking $\Lambda\to \infty$ first, we get $\langle uv\rangle=0$ as opposed to $
\langle uv\rangle =\epsilon/U'$ if $t\to\infty$ is taken first. As will be discussed below, this has to do with the fact that the linear shear keeps intact a \textit{local in time} reflection symmetry, for any finite time. It led to some confusion regarding the role of dissipation. Adding a finite friction such that $1/\alpha < |U'|^{-1}\Lambda$ keeps $\langle uv\rangle=0$, while 
a finite viscosity, combined with the linear shear, breaks this reflection symmetry so that generally $
\langle uv\rangle\neq 0$. Therefore, counter intuitively, the resulting impression is that viscosity plays a crucial role in making the momentum flux non zero (and thus sustaining the mean flow), as argued in\cite{kolokolov_structure_2016}. 

On the contrary, we argue that the ratio between viscosity and friction at small scales, quantified by $\Gamma$, does not determine if $
\langle uv\rangle =\epsilon/U'$ is satisfied. Rather, it is the condition that the dynamics at the forcing scale is dominated by shear, and not dissipation, that matters. 
The condition $1/\alpha < |U'|^{-1}\Lambda$, giving $\langle uv\rangle=0$ for an isotropic pumping in the case $\Gamma\leq 1$, translates in terms of the parameters of the problem to $K^{-1/3}\left(d/R_u\right)^2\sqrt{\delta} >1$. This is incompatible with the dynamics being dominated by shear $d/R_u\ll 1$. 
To leading order, one should therefore take $\alpha/U'\to 0$ first (keeping $t>1/\alpha$), followed by $\Lambda\to \infty$, i.e consider a linear shear mean flow with zero $\alpha$---which gives $
\langle uv\rangle =\epsilon/U'$ as we have already seen. For a discussion of the anomaly related to the limit $\alpha\to 0$ in the uniform linear shear model see Appendix \ref{app:anom_lin}.

In the case $\Gamma\gg1$ where viscosity is dominant at the forcing scale, the result $
\langle uv\rangle =\epsilon/U'$ was demonstrated in\cite{kolokolov_structure_2016}. 
It seems, however, that viscosity does not produce this result but rather does not alter it. In principle, the typical time for attenuation of a given mode by visocisty is scale dependant, which gives rise to non negligible corrections to some wavenumbers, even in the shear dominated regime\footnote{For example, considering the characteristic time scale for viscosity at the forcing scale, the condition $\gamma^{-1}<|U'|^{-1}\Lambda$ can be satisfied if $\Gamma$ is large enough: it requires that $\Gamma (d/R_u)^2 K^{-1}>1$ which does not contradict the other limits in Table \ref{table:1}.}.  However, since viscosity is less effective than shear at the forcing scale: $\gamma/U'=\Gamma \delta^{3/2}d/L \ll d/R_u\ll1$ it affects mainly very high wavenumbers, which are in any case suppressed in (\ref{eq:uv_iso}). This picture is confirmed by a direct estimate of the integrals involved in the computation of $\langle uv\rangle$\cite{kolokolov_structure_2016}.

\textit{Geometric interpretation}\\
It is useful to attach a geometric picture to the above discussion. We begin with the expression for the momentum flux due to the excitation of a given forcing mode, rewriting equation  (\ref{eq:Leb_uv}),
\begin{equation}
\begin{split}
\langle u_k^l(y) v_{-k}^{-l}(y)\rangle  =
2\eta_{k,l}\int_{-t}^{0} d\tau \frac{-k\, l(\tau)}{q^4(\tau)} .
\end{split}
\label{eq:uv_modes}
\end{equation} 
We have denoted $l(\tau)=l-U'|\tau|k$ and $q^2(\tau)=l^2(\tau)+k^2$ to make the interpretation of the integral transparent: it is the momentum flux from all modes that were excited at time $|\tau|$ in the past with wavenumber $(k,l)$, so that, due to the shear, at time $t$ they have the wavenumber $(k,l(\tau))$. At the moment of excitation, all such modes lie on a circle $l^2+k^2=q^2\approx q_f^2$.  

The main physical point is that, a linear shear tends to align the modes it acts on with its direction. Therefore, if one considers modes that were excited at far enough times in the past, only the modes very close to $k=0$ can end up in the orthogonal direction. Thus, at long enough times, a system with a cutoff around $k=0$ will have no modes in the orthogonal direction, as apposed to a system without a cutoff. The result $\langle uv\rangle=\epsilon/U'$ is a consequence of the former system, where the mean flow cannot be taken to be a linear shear for all modes. This clarifies why friction can have such a dramatic influence: it removes modes that were excited at times earlier than $\alpha^{-1}$ in the past, which, if $\alpha^{-1}$  is small enough, will erase this drastic shearing effect.      

\begin{figure*}
    \includegraphics[width=0.7\textwidth]{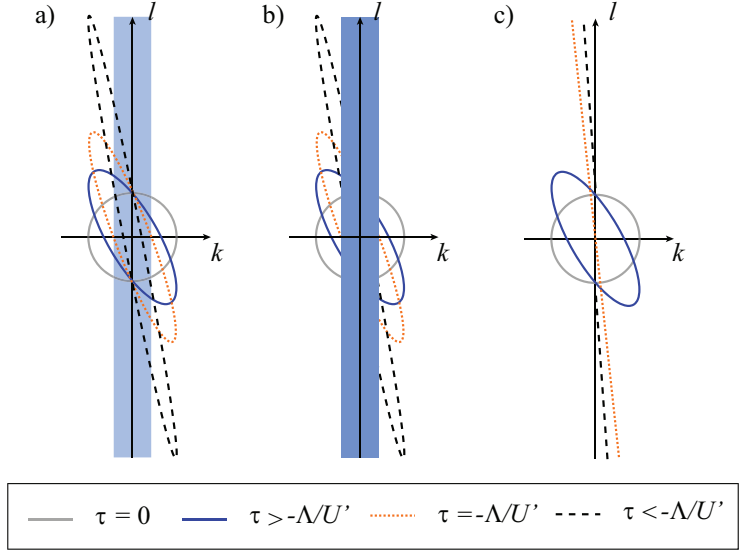}
    \centering
    \caption{Breaking of reflection symmetry by the presence of a cutoff at $k=\pm 2\pi/d$. It is assumed that $U'>0$. The shaded region represents the wavenumbers which lie outside the linear shear approximation. The solid-line circle represents the modes forced by an isotropic forcing at time $\tau=0$. The solid-line, orange dotted-line and black dashed-line ellipses correspond to modes excited at time $\tau>-U'^{-1}\Lambda$,$\tau=-U'^{-1}\Lambda$ and $\tau<-U'^{-1}\Lambda$ respectively. a) The effect of a linear shear on all excited modes. b) Modes lying outside the linear shear approximation are removed.  c) The limit $\Lambda\to \infty$ is taken in b), collapsing the dotted-line and dashed-line ellipses into tilted lines that are asymptotic to the $l$-axis. For $\tau\leq-U'^{-1}\Lambda$ the reflection symmetry is broken, as modes are absent from the first and third quadrants.}
    \label{fig:1}
\end{figure*}

Focusing on an isotropic forcing such that modes on the circle $l^2+k^2=q^2\approx q_f^2$ are excited with equal amplitude, let us explain in detail why linear shear preserves a reflection symmetry at any finite time  (which keeps $\langle uv\rangle=0$), and how it is broken when a cutoff is introduced.
In the absence of an IR cutoff for $k$, modes excited at any finite time $\tau$ in the past, are deformed such that they span a tilted ellipse at time $t$, parameterized by $l(\tau)=l_0+U'\tau k_0$, $k(\tau)=k_0$, see Figure \ref{fig:1}~a).
Any such ellipse has two separate reflection symmetries relating modes with the same $q(\tau)$: reflection with respect to the major and minor axis . These separate reflection symmetries for each time $\tau$, combined with the forcing being isotropic, is what makes the Reynolds stress zero. Now, let us introduce a cutoff for $k$ which eliminates $|k|\ll 2\pi/d$, removing a number of modes of the order of $\Lambda^{-1}\to 0$, Figure \ref{fig:1}~b). Consider the modes at the cutoff, i.e $|k|=2\pi/d$ (so that $l_0\approx q_f$), which were excited at time $\tau$. If $|\tau|< U'^{-1}\Lambda$ then for these modes $l(\tau) \neq 0$ and in the limit $\Lambda \to \infty$ (here equivalently $d\to \infty$) all modes excited at time $\tau$ still form a closed ellipse. On the other hand, if $|\tau|= U'^{-1}\Lambda$ then the modes at the cutoff have $(k,l(\tau))=(2\pi/d,0)$ at time $t$, and the limit $d\to \infty$ collapses the ellipse to a tilted line. Ellipses originating further back in the past $|\tau|> U'^{-1}\Lambda$, are split into two lines, asymptotically aligned with the $l$ axis at $\tau\to -\infty$, in this limit. They remain however asymptotically tilted, pointing at $\theta=\pi/2^{+}$ and $\theta=-\pi/2^{+}$ for $U'>0$, See Figure \ref{fig:1}~c).  
Thus, the reflection symmetry is broken for $|\tau|>U'^{-1}\Lambda$ and $\langle uv\rangle \neq 0$ is possible.

Now lets consider the effect of friction and viscosity. A finite friction, $1/\alpha < U'^{-1}\Lambda$, is effectively the same as considering a finite time $T< U'^{-1}\Lambda$, as it uniformly eliminates contributions from distant enough times in the past. Thus if $1/\alpha < U'^{-1}\Lambda$ the local in time reflection symmetry remains intact. Viscosity, on the other hand, does break the reflection symmetry at any time. It eliminates modes based on the magnitude of the radius $q(t)$ during the entire evolution. As the pairs of modes that are related to each other by reflection at the final time, have a different radius at any former time, they are suppressed differently by viscosity\footnote{ 
The former claim is evident if one considers the points $(0,q_0),(0,-q_0)$ which are invariant under the dynamics. Moreover, these are the only points whose distance from the origin is conserved. For the tilted ellipses these two points are related to each other by the $l\to -l,k\to -k$ symmetry, rather than a single reflection symmetry, meaning that their partners under reflection, at any given time $t$, necessarily had a different radius  $q(t)\neq q_0$ at any other time $t$}.  However, as discussed above, the effect of viscosity compared to that of the shear is negligible in the shear dominated regime. Thus, a small but finite viscosity effectively serves as a regularization for the integrals, which chooses the symmetry breaking answer for the Reynolds stress $\langle uv\rangle =\epsilon/U'$ without altering it, independently of the order of integration.

\textit{An inhomogeneous pumping}\\
Let us discuss the effect of an inhomogeneous pumping on the Reynolds stress. 
Generally, we expect that if the spatial variation of the forcing is on a scale much larger than its correlation length, then the same relation as for the homogeneous case would be satisfied locally: $\langle uv\rangle = \epsilon(y)/U'$. The momentum flux can, however, be very different for a variation on a scale smaller, but comparable to, the correlation length.

We focus on a jet-like mean flow with a mean velocity $U(y)$, and a $y$-dependent (but not $x$) enstrophy injection rate. As previously, we put both viscosity and friction to zero. For a homogeneous pumping, their presence gives higher order corrections, and we expect this also to be the case for an inhomogeneous pumping (we will provide some justification below).  
Denoting the Fourier coefficient of the enstrophy covariance by $\langle g_{kl}(t)\cdot g_{k'l'}(t')\rangle\equiv 2\eta_{k,ll'}2\pi \delta(k+k')\delta(t-t')$ a direct computation gives the general formula (see Appendix \ref{sec:appB}),
\begin{equation}
\langle uv\rangle =  \int \frac{dl'dldk }{(2\pi)^3}\frac{ e^{iy(l+l')}\eta_{k,ll'}}{U'(y)}\frac{\arctan(l/k)+\arctan(l'/k)}{(l+l')k}
\label{eq:uv_inh}
\end{equation}

To see the implications of this formula, it is useful to consider a specific model for an inhomogeneous pumping in the $y$ direction (details of the calculations can be found in Appendix \ref{sec:appB2}). We take the vorticity covariance to be of the form 
\begin{equation}
\begin{split}
\langle g(\bm{x})g(\bm{x'})\rangle= 2\bar{\eta}\cos sy \cos sy' J_0(q_f r)2\delta(t+t')= \\
=\bar{\eta}(\cos s(y+y')+ \cos s(y-y')) J_0(q_f r)2\delta(t+t')
\end{split}
\end{equation}
where $r^2=(x-x')^2+(y-y')^2$, $J_0$ is the Bessel function of order zero and $s>0$. For $s=0$ the forcing is both homogeneous and isotropic with correlation length equal to $q_f$.  
It is simplest to consider the limit $s \gg q_f$, so that the characteristic gradient of $\eta(y)=\bar{\eta}(1+\cos 2sy)$ is much larger than $q_f$. The energy injection rate can be shown to be 
 \begin{equation}
 \epsilon(y)=\frac{\bar{\eta}}{s^2-q_f^2}- \cos 2sy\frac{\bar{\eta}}{s^2+q_f^2}\approx \bar{\epsilon}(1-\cos 2sy).
\end{equation}  
denoting $\bar{\eta}/s^2\equiv\bar{\epsilon}=1/L \int dy \epsilon(y)$.
A direct computation of (\ref{eq:uv_inh}) then gives the Reynolds stress
\begin{equation}
\langle uv\rangle\approx\frac{\bar{\epsilon}}{U'}\left(1+\frac{s}{q_f}\log\Lambda\cos 2sy\right) =\frac{\bar{\epsilon}}{U'}-\frac{\delta \epsilon(y)}{U'}\frac{s}{q_f}\log\Lambda
\label{eq:uv_cos}
\end{equation}
to leading order, with $\delta \epsilon(y)=\epsilon(y)-\bar{\epsilon}=-\bar{\epsilon}\cos2sy$. There are a few generic features which are worth noting in this result. First, the Reynolds stress is dominated by $\delta \epsilon(y)$ (since $\Lambda \gg1$), the inhomogeneous part of the energy injection rate. Furthermore, it comes with a minus sign compared to $U'$. Thus, the variations in the energy injection rate are strongly amplified, creating large fluxes of energy (i.e $\langle vp\rangle\neq 0$ ) from maximum to minimum . These fluxes, which outweigh the locally injected energy, are compensated by a local transfer of energy from the fluctuations to the mean flow at a minimum ($U'\langle uv\rangle>0$), and from the mean flow to fluctuations at a maximum ($U'\langle uv\rangle<0$). So, while the contribution of $\delta \epsilon(y)$ averages to zero globally, and on balance energy goes from the fluctuations to the mean flow, the local picture is wildly different. 
The logarithimic dependence on $\Lambda$ is another striking feature of equation (\ref{eq:uv_cos}).

The sign of the inhomogeneous contribution and its logarithmic dependence on $\Lambda$ are both features that carry over to $q_f\lesssim s$.
Indeed, the minus sign is a consequence of the fact that $\text{sign}[\epsilon_{k,ll'}]= -\text{sign}[\eta_{k,ll'}]$ for $s>q_f$, see also Appendix \ref{sec:appB1}.
Moreover, it is apparent from expression (\ref{eq:uv_inh}) that the contribution of modes with $k\approx 0$ to $\langle uv\rangle$, which gives rise to the logarithmic term in (\ref{eq:uv_cos}), very much depends on the sign of $s-q_f$\footnote{In the limit $s\ll q_f$ one has  \unexpanded{$[\arctan((s+j)/k)+\arctan((s-j)/k)]\to 2sk/(j^2+k^2)$}, and the local relation \unexpanded{$\langle uv\rangle = \epsilon(y)/U'$}}. Let us parameterize in (\ref{eq:uv_inh}) $l=s+j$ and $l'=s-j$ such that $ j^2+k^2=q_f^2$ and $l+l'=2s$. If $s\ll q_f$  then $[\arctan((s+j)/k)+\arctan((s-j)/k)]\propto k$ while for $s>q_f>j$,   $[\arctan((s+j)/k)+\arctan((s-j)/k)]\propto \text{const}$. Thus, for $s>q_f$, modes with $k\approx 0$ provide the overwhelmingly dominant contribution, which scales like $\log \Lambda$. For $s<q_f$  on the other hand, a logarithmic term is absent. 

We do not analyze in detail the influence of dissipation on the above picture. However, on general grounds, we expect a small enough friction or viscosity to essentially leave it intact. Indeed, the logarithmic contribution comes from modes with a wavenumber no larger than $s$---the forcing wavenumber (starting at $s$ and decreasing during the dynamics). We have assumed throughout, however, that dissipation does not have a large effect on modes at the forcing scale. For friction, the condition to be able to neglect it is given by $1/\alpha > (s/q_f) \Lambda/U'$ . This estimate comes from the requirement that modes at the cutoff $k\approx 2\pi/d$ have enough time to reach $l(\tau)=0$ (see the discussion in Appendix \ref{sec:appB1} on the time $\tau$ that gives the main contribution).  
 
\subsection{Region of validity of the vortex profile}
\label{sec:validity}
Let us return to the vortex geometry, where the mean velocity is characterized by the radius $r$. At a given point, the characteristic scale of the mean velocity is of the order of the radius, meaning that $\Lambda = rq_f$. 
The approximation used to derive the profile (\ref{eq:vortex_sol}) breaks down both at small $r$ and large $r$. At small $r$, it is clear that the angular velocity $U/r$ cannot continue up to the vortex core. Indeed, we have seen in section \ref{sec:uvU} that the approximation breaks before that, at $\Lambda=rq_f\approx 1$. If $\Gamma\gg1$ then, for the mean flow, viscosity would become comparable to friction before the forcing scale is reached. The core would be determined by viscosity at the scale $\sim (\nu/\alpha)^{1/2p}= \Gamma^{1/2p}K^{-1}L$. We have also seen that $r$ cannot be too large, $r\ll R_u$ is required to have a large enough shear rate.  
In summary, the profile (\ref{eq:vortex_sol}) is expected to hold in the region $R_c\ll r\ll R_u$ with $R_c=\text{max}(K^{-1}L,\Gamma^{1/2p}K^{-1}L)$\cite{kolokolov_structure_2016}.

Of course, one may have $R_u\gg L$ in which case the size of the vortex should be determined by other considerations. One such consideration is global balances.
Indeed, the region outside the vortex plays an important role in the energy balance. 
The profile (\ref{eq:vortex_sol}) implies a non zero flux divergence of (mean) energy: 
$
r^{-1}\partial_r \left[r \langle uv\rangle U\right]=-2\epsilon.
$
Thus, most of the energy dissipation by the vortex (equal to $3\epsilon$ at a each point) is due to an energy flow from a region exterior to the vortex, rather than being injected locally. Since the vortex core region is small, this energy must come from large radii that lie outside the vortex solution, and which play the role of an energy source. 
To balance energy injection and dissipation, the area corresponding to the profile (\ref{eq:vortex_sol}) should therefore be smaller than $1/3$ of the box area, for a box with rigid walls, containing a single vortex. For the dipole, the area should be less than $1/6$ for each vortex. 
Also note that the enstrophy flux in the  region $R_c\ll r\ll R_u$ is equal to $1/r \partial_r \left(r \Omega \langle v \omega \rangle\right)=-\alpha \Omega^2+\langle v \omega \rangle \partial_r \Omega=-\epsilon/r^2$, implying that the outer region serves also as an enstrophy source for the mean flow.
\subsection{Jets in a periodic box}
\label{sec:jets}
\begin{figure}
\includegraphics[width=0.8\linewidth]{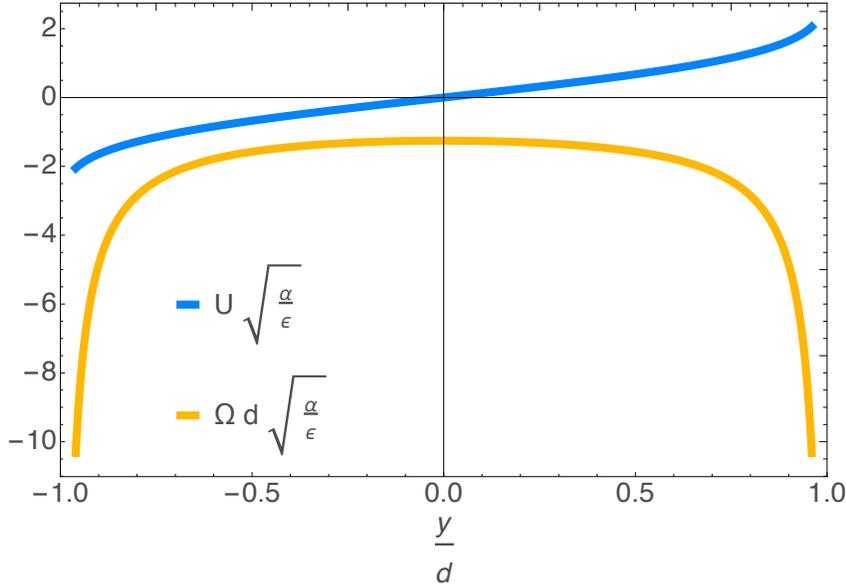}
\centering
\caption{The jet profile, equation (\ref{eq:jet}).}
\label{fig:jet}
\end{figure}

The mean velocity profile for a vortex, the solution of equations (\ref{eq:momentum}),(\ref{eq:uv}), was presented in section \ref{sec:2}. Here we consider the solution for a jet mean flow. Two opposite facing jets are expected to emerge in a rectangle, doubly periodic box with aspect ratio different from unity \cite{bouchet_random_2009}, for a homogeneous, zero-mean forcing. 
We denote by $U(y)$ the mean velocity corresponding to the jets, assumed to point in the $x$ direction. The fluctuations are assumed to be statistically independent of $x$. 
In the regime of Table \ref{table:1}, we can follow the reasoning of Sec. \ref{sec:uvU} for the vortex, and arrive at the equations for the jet mean flow away from the jets maxima where $U'=0$, see also\cite{falkovich_interaction_2016}
\begin{align}
\partial_y \langle uv\rangle=-\alpha U && U'\langle uv\rangle =\epsilon. 
\label{eq:jet_eq}
\end{align}  
As generally noted previously\cite{woillez_first_2016}, these equations can be integrated directly. Eliminating $\langle uv\rangle$: $(\ln U')'=(\alpha/\epsilon U^2/2)'$, the solution in the region $-L/4<y<L/4$ is
\begin{align}
U(y)= \sqrt{\frac{2\epsilon}{\alpha}} \text{InverseErf}\left[\frac{y}{d}\right] \\
\langle uv\rangle =d \sqrt{\epsilon \alpha} \sqrt{\frac{2}{\pi} }e^{-\left(\text{InverseErf}\left[\frac{y}{d}\right]\right)^2}
\label{eq:jet}
\end{align}
where $\text{InverseErf}$ is the inverse of $\text{Erf}(z)=2/\sqrt{\pi}\int_0^z e^{-u^2}du$, we have chosen $U(0)=0$ and $d$ is a free parameter corresponding to a characteristic length scale. In the regime of Table \ref{table:1} we expect $d\propto L/2$. Note that the solution diverges at $y=\pm d$, so it can describe only the region $-d<y<d$. The jet profile and the corresponding mean vorticity for this region are plotted in Fig.~\ref{fig:jet}. The solution for the region $L/4<y<3L/4=-L/4$ is obtained by reflection around $y=L/4$. For convenience, we will focus on the region $-L/4<y<L/4$ in the following. 

Where would this solution be applicable? The approximation breaks down either if the shear is not strong enough, i.e $(U/U')/R_u\ll 1$ is not satisfied, or if the mean flow cannt be approximated by a uniform shear---$(U/U')q_f\gg1$ is not satisfied. There can also be a region where viscosity should be taken into account for the mean flow. 
The point $y= L/4$  defines the jets maximum, where $U'=0$\footnote{In the region of the maximum, the energy flux should give the main contribution}, so that naively we would expect the shear rate to increase as one moves away from this point, towards $y=0$ (where $U=0$), and the approximation to breakdown somewhere in between---the shear being too weak. On the contrary, the profile (\ref{eq:jet}) has $|U'|$ increasing with the distance from $y=0$. This is analogous to the vortex, where the shear increased towards the center of the vortex, and originates from a symmetry in both cases. Indeed, the system is symmetric with respect to reflection around the the maximum of $|U'|$. Therefore, the momentum flux $\langle uv\rangle$ must be zero at this point. It follows that to satisfy (\ref{eq:jet_eq}) the shear rate must diverge there, and in particular it increases as one comes closer to the maximum. Similarly, for the vortex the angular momentum at the center of the vortex must vanish, implying that the angular velocity gradient increases. For the jet, this gives a  region of order $R_c=\text{max}(K^{-1}L,\Gamma^{1/2p}K^{-1}L)$ around $y=\pm d$, where (\ref{eq:jet_eq}) is not applicable. 
To sum up, (\ref{eq:jet}) is applicable in the region $-d+R_c < y<d- R_c$, with $d\ll R_u$ and $ U/d$ is the characteristic mean shear rate in most of this region.

\begin{figure}
\includegraphics[width=\linewidth]{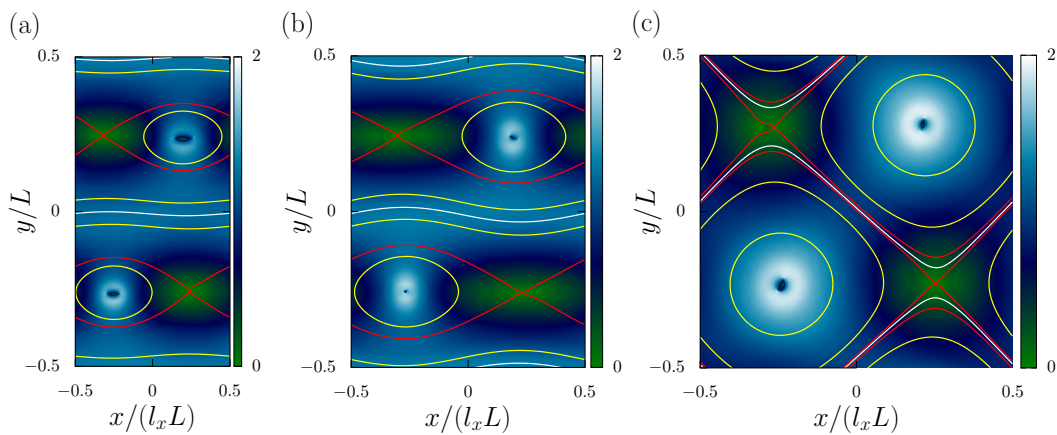}
\caption{Heat maps of the scaled speed ($\sqrt{\alpha/\epsilon} |\bf{v}|$) averaged over time $\tau_m$: (a) domain aspect ratio $1/2$ (b) domain aspect ratio $3/4$, and (c) square box. Overlaid are streamlines, red lines are separatrices. Reprinted figure with permission from\cite{frishman_jets_2017} \copyright (2017) by the American Physical Society}
\label{fig:jetsvor}
\end{figure}

\textit{Comparison to DNS:} \\
Direct numerical simulations in a periodic box with aspect ratio different from unity were performed in\cite{frishman_jets_2017}. Surprisingly, in addition to jets, large scale coherent vortices were observed, Figure \ref{fig:jetsvor}. The number of vortices and their relative motion, as well as the number of jets, was observed to change with the domain aspect ratio. When averaged over times comparable to $\tau_m$, the vortices remained pinned and the mean flow was a mixture of jets and vortices. 
It appears that the above described theory cannot be used to predict these features, its starting point being a symmetry of the mean flow that seems to be broken. Indeed, the form $U(y)$ was deduced from the symmetries of the domain, but we are not guaranteed that they remain intact. Averaging over long enough times, we do however expect all symmetries to be restored. 

\begin{figure}
\includegraphics[width=\linewidth]{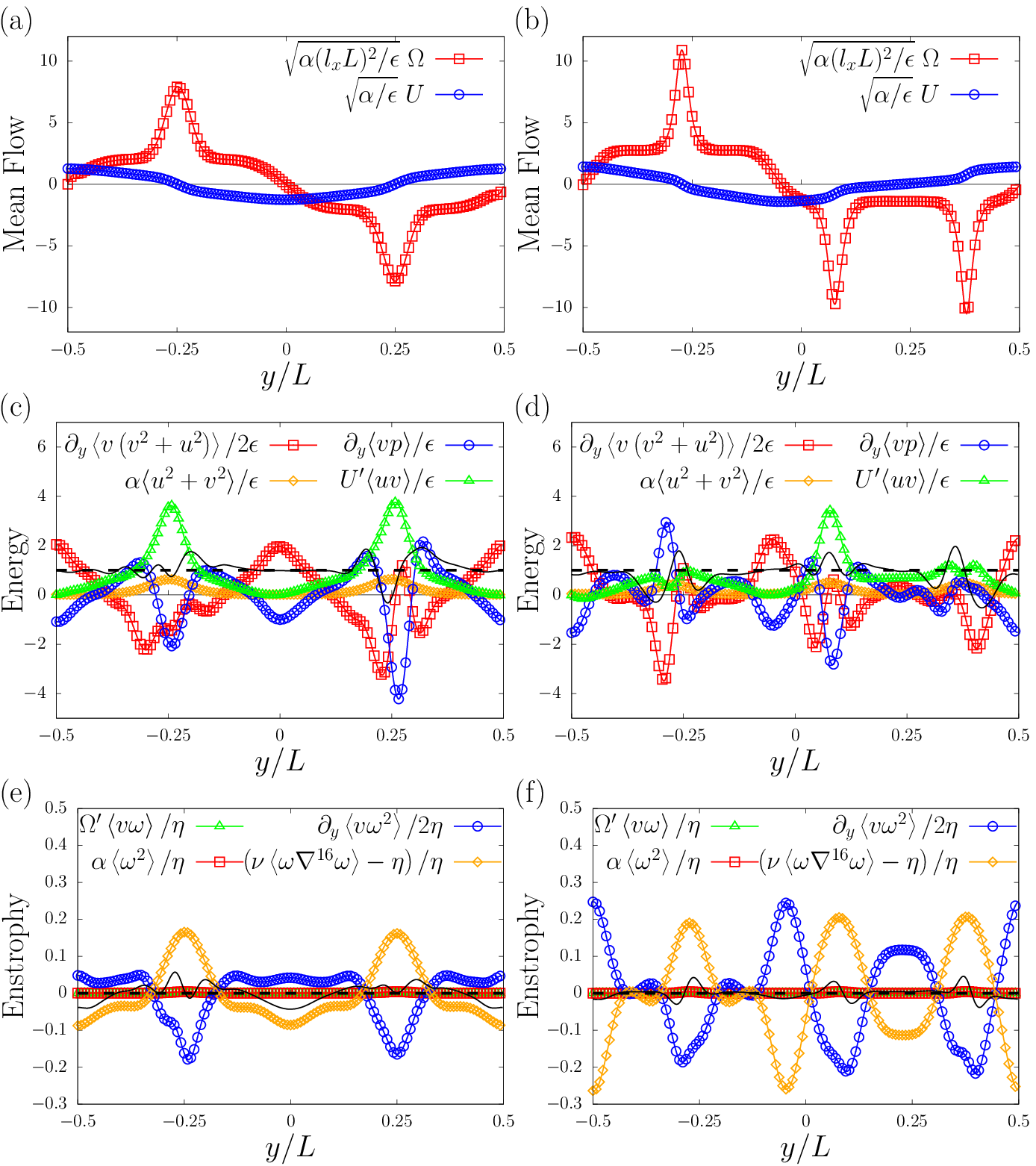}
\caption{Mean profiles of vorticity and velocity (top), balances of energy (middle) and enstrophy (bottom) for run A (right) and run F (left). Horizontal dashed lines indicate the expected  balance, solid lines indicate the numerical sum. Data were averaged over the simulation time (much longer than $\tau_m$). Reprinted figure with permission from\cite{frishman_jets_2017} \copyright (2017) by the American Physical Society.}
\label{fig:balance}
\end{figure}
It was found in\cite{frishman_jets_2017}, that when averaged up to $\tau_m \ll t\leq 1/\alpha$, homogeneity in the $x$ direction is restored, due to vortices motion, while  translational invariance in $y$ remains broken. This gives a mean flow of the form $U(y)$, but the quasi-linear approximation obviously fails at the location of the vortices. There, the mean flow and fluctuations, both of which have contributions from vortices, are of the same order.

Nonetheless, the theory knows nothing about the vortices, so if we are in the right regime it should be applicable. For aspect ratio $1/2$, which will be our focus here, the range of parameters used in\cite{frishman_jets_2017} is as follows: $\Gamma \sim O(1)$, $\delta \sim O(10^{-3})$ (note that our definition of $\delta$ differs from that of \cite{frishman_jets_2017} by a factor of $(2\pi)^{-2/3}$) and  $K=10^2$, giving $R_u/L\sim O(1)$. There were four simulations in total. The two with largest $\delta$ found two opposite-signed vortices, each vortex placed at a zero of the mean velocity.  The two other simulations, with smaller $\delta$, had three vortices, with two same-signed vortices placed on the two sides of a mean velocity zero, and the third vortex located at the second zero, Figure \ref{fig:balance} a). The transition between two and three vortices happened between $\delta= 1.6 \times 10^{-3}$ and $ 8.1 \times 10^{-4}$, with $R_u/L$ changing from $1.14$ to $1.6$. 

All four simulations correspond to the asymptotic regime $K\gg1$, $\delta \ll1$, with $\Gamma$ satisfying the requirements of Table \ref{table:1}. As all simulations had the same $K$, but decreasing $\delta$, it seems plausible that the transition in the number of vortices corresponds to an interchange of the order of limits $\delta \to 0$ and $K\to \infty$. A measure of the order of limits is the ratio $R_u/L$ which was of order unity, and which grows with decreasing $\delta$.

Can we match solution (\ref{eq:jet}) to the velocity profile observed in simulations?
This solution corresponds to the limit $R_u/L \gtrsim 1$, i.e presumably the simulations with three vortices, since the mean velocity gradient must satisfy $l_f \ll d \ll R_u$.  
Indeed, if $R_u/L$ is too small then so is the parameter $d$ in (\ref{eq:jet}), and the region where this solution can be applied vanishes. The profile  (\ref{eq:jet}) seems compatible with the one realized between the two same-signed vortices in the simulations with $R_u/L\geq 1.6$: the mean vorticity being almost constant there, see Figure \ref{fig:balance} a), b) and compare to Figure \ref{fig:jet}. Of course, a quantitative comparison with (\ref{eq:jet}) should be done to confirm this hypothesis, determining the parameter $d$ from $U'(0)$ in the simulations. 

Note that contributions coming from non-linear terms have been found to be significant for the enstrophy balance, see Figure \ref{fig:jetsvor} c). This was the case even in the region between the same-signed vortices, where the profile  (\ref{eq:jet}) may apply.
If we adapt the condition under which non-linear terms can be neglected for the enstrophy, from our considerations for the vortex (Appendix \ref{app:non-linear}), we obtain the condition $d^2/L^2 \ll R_u/L K^{-1}$. Thus, even if $d/L \ll R_u/L $ as required for (\ref{eq:jet}), it is still compatible with $d^2/L^2 \ll R_u/L K^{-1}$ not being satisfied. 

It is worth commenting that the profile (\ref{eq:jet}) would not be sufficient to explain the observations in the simulations, even if it could be matched to the region between the two same-signed vortices. Most importantly, a fundamental explanation for the presence of vortices is lacking.
In addition, it remains unclear why reflection symmetry around the jets extrema is broken, so that (\ref{eq:jet}) cannot be realized around the second zero of the mean velocity (in particular having four instead of three vortices). It could just be that the simulations are in an intermediate regime, $R_u/L$ not being large enough. 
In addition to the possible appearance of one more vortex, as $R_u/L$ is increased, the magnitude of $d$ and the extent of the region of validity of (\ref{eq:jet}) can also increase. Then, if there is no global constraint restricting this region, the vortices should be squeezed closer and closer to the jets maxima. This scenario does not appear to be very likely, and it would be interesting to test what actually happens in the regime $R_u/L\gg1$. 

\section{Fluctuations}
\label{sec:fluct}
The next natural step, after the mean velocity profile is established, is to consider the average energy in turbulent fluctuations $\langle u^2\rangle$ and $\langle v^2\rangle$ (i.e the diagonal terms in the Reynolds stress tensor). We will consider the vortex geometry in this section, drawing confidence from numerical simulations that the profile (\ref{eq:vortex_sol}) is indeed realized\cite{laurie_universal_2014}.
The energy fluctuations can be obtained from the velocity covariance. The latter will be the main focus of this section, and we will not discuss $\langle u^2\rangle$ and $\langle v^2\rangle$ directly.

First, a comment on the role of the inverse cascade for velocity correlation functions is in order. In the region where (\ref{eq:vortex_sol}) applies, the dynamics is dominated by the mean flow down to the forcing scale. This implies that the inverse cascade is suppressed there. Accordingly, we do not expect third order velocity correlation functions to satisfy the usual energy flux law\cite{bernard_three-point_1999}. 

For large radii, outside the region where (\ref{eq:vortex_sol}) applies, on the other hand, the inverse cascade may be present\cite{kolokolov_structure_2016}. For two-point correlation functions, the characteristic scale for the fluctuating velocity difference is given by the distance between the two points.
Following\cite{xia_turbulence-condensate_2008} we can then define a wavenumber $k_t$, corresponding to the separation where the effects of the mean shear and the non-linear interactions are of the same order. If we assume that the characteristic scale for the mean velocity at  $r\gg R_u$ is of order $L$ then we get that $k_t=(R_u/L)^{3/2} q_f$. The inverse cascade disappears completly if $k_t>q_f$, i.e $R_u\gg L$. In\cite{xia_spectrally_2009} the third order velocity correlation function was found to scale linearly with the separation, when averaged over the entire domain, even for the strongest condensate considered. It seems that $k_t$ and $q_f$ were roughly of the same order. This may indicate that the main contribution to the third order correlation function came from regions outside the vortex, dominated by the inverse cascade.   

Let us now return to the region of the vortex corresponding to equation (\ref{eq:vortex_sol}). 
We will argue that the velocity covariance, in the main order, is insensitive to the forcing, and consequently the linear shear approximation cannot be used to determine it. Instead, it is determined by the homogeneous solutions of an advection equation, so that the detailed profile of the mean flow becomes important.

For the vortex geometry, it is natural to desompose the velocity covariance into angular harmonics. We will briefly describe how a closed equation for these angular harmonics can be derived, and discuss its solutions. We leave the full details of the derivation and an in depth discussion of the solutions to a future publication\cite{herbert_fluctuations_2017}. 

The starting point of our derivation is the $(r,r)$ component of the steady state equation $\partial_t \langle \bm{v_1}\bm{v_2}\rangle-\langle \bm{v_1}\rangle \partial_t \langle\bm{v_2}\rangle-\langle \bm{v_2}\rangle \partial_t \langle\bm{v_1}\rangle=0$.
The resulting equation contains the pressure as well as cubic velocity correlation functions. We then do the following steps: (i) We act with the operator $r_1^2 \nabla_1^2 r_2^2 \nabla_2^2 r_1 r_2$, commuting the differential operators such that the Laplacian acts directly on the pressure. (ii) We neglect the cubic terms and use that to leading order  $-\nabla^2 p= -\frac{2U}{r}\partial_r u$, for the mean flow (\ref{eq:vortex_sol}). (iii) We use incompressibility:  $\partial_{\phi_1}\langle v_2  u_1\rangle=-\partial_{r_1} r_1\langle v_2  v_1\rangle$. At this point we can take the Fourier transform in $\Delta\phi$ of the resulting equation. Indeed, the angular harmonics are defined by $
\langle v_1 v_2 \rangle= \sum_{m=-\infty}^{\infty} \langle \hat{v}_m(r_1)\hat{v}_{-m}(r_2)\rangle e^{im\Delta\phi}
$, where isotropy allows to express the correlation functions in terms of $\Delta\phi=\phi_1-\phi_2$.

Proceeding further, in the region $R_c\ll r\ll R_u$ dissipation by friction and viscosity may be neglected.
The contribution from the forcing can also be dropped, since for $r_1,r_2 \gg l_f$ it gives a nonzero contribution in a range of angles at most of the order of $\text{max}(l_f/r_1,l_f/r_2)$.    
We arrive at a scale invariant equation coming solely from the advective derivative. This prompts the use of the scaling form $\langle \hat{v}_m(r_1)\hat{v}^*_m(r_2)\rangle= r_1^{\bar{\lambda}-1}f_m(r_2/r_1)$ for the solution, allowing to convert the PDE into an ODE in the variable $R=r_2/r_1$. Finally, the resulting equation can be cast into the form of a Hypergeometric equation:
\begin{equation}
\prod_{i=1}^4\left(R\frac{d}{dR}-\gamma_i\right)f_m(R)=R\prod_{i=1}^4\left(R\frac{d}{dR}+\alpha_i\right)f_m(R)
\label{eq:hyper}
\end{equation}
where $(\gamma_1,\gamma_2,\gamma_3,\gamma_4)=(\bar{\lambda}-|m|,\bar{\lambda}+|m|,-1+\sqrt{m^2-1},-1-\sqrt{m^2-1})$, $(\alpha_1,\alpha_2,\alpha_3,\alpha_4)=(1-|m|,1+|m|,-\bar{\lambda}+\sqrt{m^2-1},-\bar{\lambda}-\sqrt{m^2-1})$. There are four families of solutions to this equation, parametrised by $\bar{\lambda}$. Choosing the relevant solutions requires additional information, which our theoretical framework lacks. In\cite{herbert_fluctuations_2017} it is supplemented with extensive numerical simulations, and information about the realized solutions is successfully extracted.  Such solutions may depend on the boundary conditions of the domain, let us therefore broadly describe the families of solutions of (\ref{eq:hyper}).
They are of the general form
\begin{equation}
\langle \hat{v}_m(r_1)\hat{v}_m^*(r_2)\rangle=  \begin{cases}  r_1^{\lambda}\left(\frac{r_2}{r_1}\right)^{\gamma_s} \sum\limits_{n=0}^{\infty} a^{s}_{|m|,\lambda} \left(\frac{r_2}{r_1}\right)^{n} \,\,  r_2<r_1\\
 r_2^{\lambda}\left(\frac{r_1}{r_2}\right)^{\gamma_s} \sum\limits_{n=0}^{\infty} a^{s}_{|m|,\lambda} \left(\frac{r_1}{r_2}\right)^{n}\,\, r_1<r_2 \end{cases}
\end{equation}
for $\gamma_i$ with $i=2,3,4$ and
\begin{equation}
\langle \hat{v}_m(r_1)\hat{v}_m^*(r_2)\rangle=  \begin{cases}  r_1^{\lambda}\left[\ln\left(\frac{r_2}{r_1}\right) \left(\frac{r_2}{r_1}\right)^{\gamma_2} \sum\limits_{n=0}^{\infty} b_{|m|,\lambda} \left(\frac{r_2}{r_1}\right)^{n}\right. \\
\left.+\left(\frac{r_2}{r_1}\right)^{\gamma_1} \sum\limits_{n=0}^{\infty} a^1_{|m|,\lambda} \left(\frac{r_2}{r_1}\right)^{n}  \right]\,\,  r_2<r_1\\
  r_2^{\lambda}\left[\ln\left(\frac{r_1}{r_2}\right) \left(\frac{r_1}{r_2}\right)^{\gamma_{2}} \sum\limits_{n=0}^{\infty} b_{|m|,\lambda} \left(\frac{r_1}{r_2}\right)^{n}\right. \\
\left.+\left(\frac{r_1}{r_2}\right)^{\gamma_1} \sum\limits_{n=0}^{\infty} a^1_{|m|,\lambda} \left(\frac{r_1}{r_2}\right)^{n}  \right]\,\, r_1<r_2 \end{cases}
\label{eq:fluc_deg}
\end{equation}
for $\gamma_1$, where $\lambda=\bar{\lambda}-1$. We do not specify the coefficients $a_{|m|,\lambda}$, which in the non-degenerate cases correspond to the coefficients of a Hypergeometric function $_4F_3$. Depending on $\bar{\lambda}$, degeneracies may occur for solutions $2,3,4$, so that they can take a form similar to  (\ref{eq:fluc_deg}). The values of $\bar{\lambda}$ are limited by the requirement that the solution $\langle \hat{v}_m(r_1)\hat{v}^*_m(r_2)\rangle= r_1^{\lambda}f_m(r_2/r_1)$ satisfies the Cauchy-Schwarz inequality $|\langle \hat{v}_m(r_1)\hat{v}^*_m(r_2)\rangle| \leq\sqrt{ \langle |\hat{v}_m(r_1)|^2\rangle \langle |\hat{v}_m(r_2)|^2\rangle} $ for any $R$ and in particular for $R\to 0,\infty$. We thus get the condition that if the dominant power law in the solution is given by solution $(1)$, then $\bar{\lambda}\geq 2|m|-1$; for solution $ (2)$: $\bar{\lambda}\geq -2|m|-1$;  solution $(3)$: $\bar{\lambda}\leq -1+2\sqrt{m^2-1}$; and
solution $(4)$: $\bar{\lambda}\leq -1-2\sqrt{m^2-1}$. 
Another property of the solutoins is that, except for particular values of $\bar{\lambda}$ for which the solutions are polynomial in $R$, the vorticity covariance generally has the behavior $\langle \hat{\omega}_m(r_1)\hat{\omega}_m^*(r_2)\rangle\propto |r_1-r_2|^{-1}$ as  $|r_1-r_2|\to l_f$.

The mode $|m|=1$ is an especially degenerate case. Two of its solutions are particularly simple. One given by    
\begin{equation}
\langle \hat{v}_{1}(r_1) \hat{v}^*_{1}(r_2)\rangle= 
\begin{cases}
 r_2^{\bar{\lambda}-1}((1-\bar{\lambda})\frac{r_2}{r_1}+1+\bar{\lambda}), & r_2<r_1 \\
 r_1^{\bar{\lambda}-1}\left((1-\bar{\lambda})\frac{r_1}{r_2}+1+\bar{\lambda}\right) , & r_1<r_2 
  \end{cases}
  \end{equation}
for $\bar{\lambda}\geq 1$, and the second solution obtained by interchanging $r_1 \leftrightarrow r_2$ in the above expression, with the requirement $\bar{\lambda}\leq-1$. 

The zeroth mode, $m=0$, deserves a separate consideration (see also\cite{kolokolov_velocity_2016,falkovich_interaction_2016}). As was noted before, and can be deduced from the $(\phi,\phi)$ component of the steady state equation $\partial_t \langle \bm{v_1}\bm{v_2}\rangle-\langle \bm{v_1}\rangle \partial_t \langle\bm{v_2}\rangle-\langle \bm{v_2}\rangle \partial_t \langle\bm{v_1}\rangle=0$, $\langle|\hat{u}_0|\rangle$ does not have a contribution from advection by the mean flow. In\cite{kolokolov_velocity_2016} it was assumed that $\langle|\hat{u}_0|\rangle$ is determined by a balance between dissipation and forcing. If that is the case, one can derive a differential equation for $\langle|\hat{u}_0|\rangle$ using that $\int_0^{2\pi} \chi_{12}^{\phi\phi} \approx \epsilon l_f/r$ for the forcing term. Note that the equation would strongly depend on the type of dissipation used, and in particular on $\Gamma$ and the power of hypervisocity $p$ (an estimate is given in \cite{kolokolov_velocity_2016}). Additionally, it seems plausible that non-linear terms would actually have a non-negligible contribution compared to dissipation terms (at least at the forcing scale). Unfortunately, it is unclear how to evaluate the former in the presence of the large scale shear, so no definite statement about $\langle|\hat{u}_0|\rangle$ can be made.

Finally, we note that the solutions to equation (\ref{eq:hyper}) are real and, since $ \langle \hat{v}_m(r)\hat{v}_m^*(r)\rangle$ is real, their combinations can only appear with real coefficients. Thus, $m\text{Re}\left[ \langle \hat{u}_m(r_1)\hat{v}_m^*(r_1) \rangle\right] =  \lim_{r_2\to r_1} \text{Im}\left[\partial_{r_1} r_1\langle \hat{v}_m(r_1)\hat{v}_m^*(r_2)\rangle\right]=0$, so that these solutions give no contribution to the momentum flux $\langle uv\rangle$, which therefore originates from a subleading contribution. This is in accordance with our expectations from symmetry considerations, as discussed in section \ref{sec:uvU}.
\section{Summary and discussion}
\label{sec:smry}
A condensate, a spontaneously forming, large coherent structure, is one of the most striking features of two dimensional turbulence. This paper has provided an overview of recent work on the structure of such condensates and the weak turbulent fluctuations around them.

The emergent strategy to derive the condensate profile, is to use the relation between the Reynolds stress and the mean shear rate $\langle uv \rangle \propto 1/U'$ (see also (\ref{eq:uv})), in conjunction with the averaged momentum balance.  For a homogeneous pumping $\langle uv \rangle=\epsilon/U'$, which can most simply be derived from the energy balance and the fact that $\langle vp \rangle=0$ due to a reflection symmetry of a linear shear flow. Most importantly, we emphasize that this result is independent of the dissipation mechanism at the forcing scale. In particular, it should apply to the case where friction dominates over viscosity, the case most frequently encountered in real life flows. 

We have presented the application of the above strategy to two simple cases, jets and vortices. It has also been recently applied to the flow on a sphere\cite{falkovich_interaction_2016}. While the theory appears to work well for the (isotropic) vortex mean flow emerging in a square box\cite{laurie_universal_2014}, some puzzels still remain in regards to its applicability to a periodic rectangular domain\cite{frishman_jets_2017}. Further study, especially via direct numerical simulations, is required to clarify this point.

An important point is that the relation $\langle uv \rangle=\epsilon(y)/U'$ holds only if $\epsilon(y)$ varies on a much larger scale than the forcing correlation length, $l_f$. A general relation can also be obtained for an arbitrary enstrophy covariance (\ref{eq:uv_inh}). It was derived in the simplest geometry---that of a jet---but it should be possible to obtain a similar expression for a different geometry, and in particular for a vortex mean flow. Interestingly, this opens the possibility to design the forcing, either in simulations or experiments, to produce a desired mean flow. For fast spatial variation of $\epsilon(y)$ (compared to $l_f$), we have seen that the spatial variations are hugely magnified in the momentum flux, causing large energy fluxes in space and a local energy exchange between mean flow and fluctuations that can be of either sign, see equation (\ref{eq:uv_cos}).        

It is worth while to discuss the relevance of the above results for two more settings that are usually of interest: a pressure driven channel flow and a flow on the beta plane (i.e differential rotation). It should be clear that the derivation above is inapplicable to a flow driven solely by pressure, as a key assumption was that the forcing is supported on small scales. The flow on the beta plane is a subtler issue. Without performing the complete analysis, it can already be stated that if $U q_f^2 \lesssim \beta$ then   
the beta effect cannot be neglected in the calculation of $\langle uv\rangle$, and since it breaks the reflection symmetry which makes $\langle vp \rangle=0$, we do not generally expect $\langle uv \rangle=\epsilon/U'$.

This paper has also presented the next step---deriving some formulas for turbulent fluctuations. 
Indeed, once the mean flow is known, it becomes possible to characterise the statistics of velocity fluctuations as well. The most salient point is the significance of zero modes for the fluctuations. For the vortex, the advection equation for the velocity covariance, the Lyapunov equation, possesses non-trivial zero modes, and the fluctuations are then determined by such modes, therefore taking a universal form. We have presented the first steps in the exploration of these modes here, their in depth investigation will be given in\cite{herbert_fluctuations_2017}.

Fifty years after Kraichnan's seminal paper, where the concept of a condensate (among other things) was first conceived\cite{kraichnan_inertial_1967}, many inexplicable features of the condensate state remain. Nonetheless, I hope to have demonstrated that 
significant progress in its understanding has been recently made; perhaps the mysteries of the condensate can be dispelled, though not its beauty.
\appendix
\begin{widetext}
\section{Comments on the correlation time of the forcing and the averaging time}
\label{app:cor_time}
\textit{The averaging time:} In a doubly periodic domain, usually used in numerical simulations, translational invariance implies that the large scale structure wonders across the domain. To avoid having a zero mean flow, for this setting our discussion will refer to averages over much shorter timescales than this slow motion.

\textit{Forcing with finite correlation time:} Let us briefly mention the influence of the correlation time of the forcing, $\tau_f$\footnote{I am grateful to S. Musacchio for an eluminating discussion of this point}. So far, we have implicitly assumed that $ \sqrt{\alpha/q_f^2 \epsilon}=K^{-1} \tau_m \gg \tau_f $, i.e that the forcing decorrelates faster than the time it takes the mean flow to sweep across a scale of the order of the forcing correlation length. In that case, the forcing is approximated to be white in time, resulting in a fixed energy injection rate. In the opposite limit, the velocity decorrelates from the forcing, due to the mean flow sweeping, during a typical time equal to $K^{-1} \tau_m$ \cite{tsang_forced-dissipative_2009}. As a result, the energy injection rate is proportional to $F^2 K^{-1} \tau_m$. It decreases with a decrease in $\alpha$ or the forcing correlation length for a fixed forcing amplitude $F$: $\epsilon\propto q_f^{-2/3}\alpha^{1/3}$, which implies $\delta \propto \alpha^{9/10} q_f^{2/9}$. 
Note that the extreme limit of a constant forcing is special, as in the steady state it injects energy only through the mean flow. To balance dissipation, the mean flow must therefore have a non zero support on the forcing modes, which is not required a priori for a finite correlated forcing. 
\section{non-linear terms and the quasi-linear approximation}
\label{app:non-linear}
To compare non-linear terms to mean flow advection we may use a naive dominant balance consideration. 
For the vortex, the characteristic scale of the mean flow gradient can be taken of the order of the radius $r$. Consider the fluctuations at the forcing scale $q_f$. Then, locally at a radius $r$, : the advection terms are $I^{v}_1\equiv U/r \partial_{\phi} v $ and $I^{v}_2 \equiv \bm v \cdot \nabla U  \sim |\bm v| U/r $, and the non-linear term is $I^{v}_3\equiv \bm{v} \cdot \nabla\bm v \sim \epsilon^{1/3}q_f^{2/3}|\bm v|$ \footnote{The pressure term $\nabla p$ contains both an advective term, $\propto I_2^{v}$, and a non-linear term $\propto I_3^{v}$.}. We can thus generally estimate the advection terms by the typical time scale related to the mean flow: $I^{v}_1\sim I^{v}_2\sim |\bm v| U/r $.
We have that $I^{v}_3/I^{v}_1 \sim I^{v}_3/I^{v}_2 =r/L \delta^{1/2} K^{2/3}=r/R_u$. Hence, we have the requirement $r/R_u\ll 1$ for non-linear interactions to be unimportant. 
f course, this estimate is not at all rigorous, since the presence of the mean flow changes the rate of non-linear transfer, and, in particular, is expected to suppress it at scales larger than the forcing scale. 

We can make a similar estimate for the vorticity: $I_1^{\omega}\equiv U/r\partial_{\phi}\omega \sim U/r \omega$, $I^{\omega}_3\equiv \bm{v} \cdot \nabla \omega \sim \epsilon^{1/3}q_f^{2/3}\omega$ and $I^{\omega}_2\equiv v \partial_r \Omega \sim U/r^2 q_f \omega$. Thus $I^{\omega}_3/ I^{\omega}_1=I^{v}_3/ I^{v}_1\ll1$ for $r/R_u\ll1$ while  $I^{\omega}_3/I^{\omega}_2\sim \Lambda \, r/R_u $.
However, if one considers the enstrophy balance at steady state the contribution of $I^{\omega}_1$ vanishes due to isotropy, and one needs $ \Lambda \, r/R_u\ll1 $ for non-linear terms to be negligible. Thus, non-linear terms become negligible in the limit $\delta\to 0$ followed by $K\to\infty$, both for the velocity and vorticity dynamics. However, for a given $K$, $\delta$ needs to be much smaller for these terms to be unimportant for the vorticity dynamics rather than for the velocity. 
\section{The shear approximation: non-commuting order of integration and the anomaly of the linear shear model}
\subsection{Validity of the shear approximation---neglecting $vU''$} 
\label{sec:app_shear}
We we would like to give a more conservative estimate for the validity of the shear approximation $|\partial_t\omega^l_k+U(y) ik\omega_k^l | \gg |-v_k^l U''(y)| $. Let us comment that the following considerations do not rely on the use of local coordinates,  i.e Cartesian coordinates, and should directly carry over to polar coordinates, relevant for the vortex. 

Recall that the notation $\omega_k^l$ corresponds to the response of the vorticity to a forcing mode with wavenumbers $(k,l)$ (which could have been excited at any time in the past). The mean flow  and forcing being homogeneous in $x$, the wavenumber $k$ remains intact throughout the dynamics of $\omega_k^l$. On the other hand, the $y$ direction wavenumber $l(t)$ can change compared to that of the initially excited mode $l$. We can estimate $v_k^l \sim \omega_k^l i k/(k^2+l(t)^2)$ with $l(t)$ the smallest wavenumber that was produced up to time $t$. This gives  $v_k^l \sim \omega_k^l i k^{-1}$ if we assume the dynamics can produce arbitrarily small $l(t)$ (in practice this is not necessarily true for all $k$). Therefore a more conservative condition for $|\partial_t\omega^l_k+U(y) ik\omega_k^l | \gg |-v_k^l U''(y)| $ to hold becomes $kd\gg1$. The latter condition also arises at a later stage in the derivation in the main text, so initially using the loose estimate $q_fd\gg1$ does not change the subsequent results.

\subsection{Non-commuting order of integration for an isotropic homogeneous pumping}
\label{sec:app_noncom}
Recall that the shear approximation is valid only if we consider the wavenumber in the local $x$ direction, $k$, that is not too small: $k d \gg 1$ where $d$ is a characteristic length scale for the mean flow. In terms of the angle $\tan \theta= l/k$, if we assume that $l^2+k^2 \approx q_f^2$, we get that angles in the sections $-\pi/2-\Lambda^{-1} <\theta< -\pi/2+\Lambda^{-1}$ and $\pi/2-\Lambda^{-1} <\theta< \pi/2+\Lambda^{-1}$ with $\Lambda = q_f d$ lie outside our approximation. Our claim is that if one sets $T \to \infty$ and $\Lambda \to \infty$ simultaneously the resulting integral expression for the momentum flux depends on the order of integration. The momentum flux for finite $\Lambda$ and $T$ is given by
\begin{equation}
\begin{split}
\left(\int_{-\pi/2+\Lambda^{-1}}^{\pi/2-\Lambda^{-1}} +\int_{\pi/2+\Lambda^{-1}}^{3\pi/2-\Lambda^{-1}}\right)\frac{d\theta}{2\pi} \int_{-T}^{0} d\tau \frac{\tan\theta +U'\tau}{(1+(\tan\theta +U'\tau)^2)^2\cos^2\theta}=\frac{1}\pi\int_{-\Lambda}^{\Lambda} dy \int_{-T}^{0} d\tau \frac{y +U'\tau}{(1+(y +U'\tau)^2)^2} 
\end{split}
\end{equation}
 Let us consider finite $T$ and $\Lambda$ and then take the limits $T\to \infty$ and $\Lambda\to \infty$ in different orders in the final result:
\begin{equation}
\begin{split}
\frac{1}\pi\int_{-\Lambda}^{\Lambda} dy \int_{-T}^{0} d\tau \frac{y +U'\tau}{(1+(y +U'\tau)^2)^2} 
=-\frac{1}{2\pi} \int_{-T}^{0} d\tau\left( \frac{1}{(1+(\Lambda +U'\tau)^2)}-\frac{1}{(1+(-\Lambda +U'\tau)^2)}\right)\\=-\frac{1}{U'2\pi}(2\arctan \Lambda -\arctan (\Lambda-U'T)-\arctan (\Lambda+U'T))
\end{split}
\end{equation}
Now, if we first take the limit $\Lambda\to \infty$ we get zero (corresponding to integrating over the angle first) while taking $T\to \infty$ first, followed by $\Lambda\to \infty$ we get $-1/(2U')$ . The main contribution to the integral comes from times $U'\tau \approx \Lambda$. Modes that were excited at this time with the smallest possible $k$: $k\approx 2\pi/d$, i.e at the IR cutoff, have zero $y$- direction wavenumber $l(t)=0$ at the measurement time due to the shear. The limit $\Lambda \to \infty$ then makes  $q(t)^2=l(t)^2+k^2 \to 0$ for these modes, collapsing the ellipse to a line that passes through the origin in Fourier space. Before this point in time the excited modes remain in an ellipse and so preserve the reflection symmetry. Modes that are excited after this point in time are gradually all pushed to an infinite radius, and so eventually cease to contribute significantly to the integral. 

In the presence of uniform friction, this picture would therefore be unchanged as long as $U'\alpha \gg \Lambda$, so that if it so happens that friction becomes a dominant term (compared to non-linear ones) then still one would have $\langle uv \rangle \neq 0$. 

We should also address the choice of $\Lambda$: it should reflect the ratio between a characteristic scale of the mean velocity and the forcing correlation length scale. If this ratio is very large, i.e $\Lambda \to \infty$ then our shear approximation is justified. In order for this argument to not be circular (since the cutoff depends on the velocity gradients which we do not know a-priori) we assume that there is some reasonable characteristic scale for the mean flow, determined by the large scale geometry. For a channel flow one would expect that ultimately $d\to L/2$ while for the vortex the local scale would be determined by the radius $r$.   
Since the end result does not depend on the cutoff we can also work directly in the limit $T\to \infty$ and $\Lambda \to \infty$, but then have to be careful about the order of integration. 

\subsection{Anomaly of the linear shear model}
\label{app:anom_lin}
Interestingly, the seemingly simple linear shear model (i.e in the absence of a cutoff) has an anomaly: the limit $\alpha\to 0$ gives a different Reynolds stress than the one obtained setting $\alpha=0$ from the start (performing the time integration first in both cases)\cite{srinivasan_reynolds_2014}. Indeed, for $\alpha=0$ the result is independent of the angular details of the forcing covariance, whereas in the limit $\alpha\to 0$ it does depend on the degree of anisotropy\cite{srinivasan_reynolds_2014}.  Note that it is not entirely clear where the latter model could be relevant (in the present context), since outside the shear dominated regime there is no reason for non-linear interactions to be suppressed. Moreover, for an isotropic forcing it is clearly necessary to include such interactions, otherwise the mean flow cannot be sustained if it is not driven externally: $\langle uv\rangle=0$ and all of the injected energy is dissipated directly by fluctuations.

\section{Calculation of the Reynolds stress for an inhomogeneous pumping}
\label{sec:appB}
We assume a white in time forcing, such that the energy injection rate $\epsilon(x,y)$ is given by 

\begin{equation}
\begin{split}
\langle \bm{f}(t)\cdot \bm{f}(t')\rangle =2 \epsilon(x,y)\delta(t-t')= 2 \delta(t-t')\int \frac{d\bm{q}}{(2\pi)^2} \int \frac{d\bm{q'}}{(2\pi)^2}   e^{i(l+l')y}e^{i(k+k')x}\epsilon_{kl,k'l'} \\ =\int \frac{d\bm{q}}{(2\pi)^2} \int \frac{d\bm{q'}}{(2\pi)^2}  \left(\langle f^x_{kl} (t)  f^x_{k'l'} (t')\rangle+\langle f^y_{kl} (t)  f^y_{k'l'} (t')\rangle\right) e^{i(l+l')y}e^{i(k+k')x} 
\end{split}
\end{equation}

The enstrophy injection rate is
\begin{equation}
\langle g(t)\cdot g(t')\rangle = \int \frac{d\bm{q}}{(2\pi)^2} \int \frac{d\bm{q'}}{(2\pi)^2}  \left(-l l'\langle f^x_{kl}   f^x_{k'l'} \rangle- k k'\langle f^y_{kl}   f^y_{k'l'} \rangle+kl'\langle f^y_{kl}  f^x_{k'l'} \rangle+k'l\langle f^x_{kl}   f^y_{k'l'} \rangle\right) e^{i(l+l')y}e^{i(k+k')x}
\end{equation}

using the assumption that the forcing preserves incompressibility we have $k f^x_{kl}+l f^y_{kl}=0$ implying that $\langle f^x_{kl}   f^x_{k'l'} \rangle =ll'/(kk')\langle f^y_{kl}   f^y_{k'l'} \rangle$ and $\langle f^x_{kl}   f^y_{k'l'} \rangle=-l/k\langle f^y_{kl}   f^y_{k'l'} \rangle$. Then, $\langle g_{kl}(t)\cdot g_{k'l'}(t')\rangle = -\bm{q}^2 \bm{q'}^2/(kk') \langle f^y_{kl}   f^y_{k'l'} \rangle$. We obtain the relation between enstrophy and energy injection rate:
\begin{equation}
\langle g_{kl}(t)\cdot g_{k'l'}(t')\rangle\equiv 2\eta_{kl,k'l'}\delta(t-t') =-\frac{2\epsilon_{kl,k'l'}\bm{q}^2 \bm{q'}^2}{kk'+ll'} \delta(t-t')
\end{equation}
We have been considering a system where the mean flow depends only on $y$, which we do not expect if the energy injection rate depends also on $x$. We will therefore consider only a homogeneous in $x$ forcing:
\begin{equation}
\begin{split}
\langle g_{kl}(t)\cdot g_{k'l'}(t')\rangle\equiv 2\eta_{k,ll'}2\pi \delta(k+k')\delta(t-t')\\ =-\frac{2\epsilon_{k,ll'}\bm{q}^2 \bm{q'}^2}{-k^2+ll'} \delta(t-t')2\pi \delta(k+k')
\end{split}
\end{equation}
Note that in our notation $\epsilon_{k,ll'}=\epsilon_{-k,l'l}$ since the forcing correlation function is symmetric with respect to the exchange of the pairs $k,l$ and $k',l'$.
We would like to solve the equation 
\begin{equation}
\partial_t \omega+U \partial_{x} \omega = g 
\label{eq:w_fluct_min_app}
\end{equation}
for an inhomogeneous pumping. Following \cite{woillez_first_2016}, we will work in Fourier space in $x$, for a given Fourier mode of the forcing in $y$: $g_{k}^l(y)=g_{kl} e^{ily}$ and $\langle g_{k}^l(y,t)g_k'^{l'}(y',t')\rangle =2\eta_{kl,k'l'} e^{i(ly+l'y')}\delta(t-t')\delta(k+k')$. The velocity is obtained from the vorticity, via the stream function $(\partial_y^2 -k^2) \psi_{k}^l(y)=\omega_{k}^l(y)$. For a given mode $k\neq 0$, the stream function can be exactly expressed in terms of the vorticity using the Greens function of the equation (see \cite{morse_methods_1946} or \cite{woillez_first_2016}):
\begin{equation}
\psi_{k}^l(y,t)= -\frac{1}{2k^2}\int_{-\infty}^{\infty}dY e^{-|Y|}\omega_{k}^l\left(y-\frac{Y}k,t\right)
\label{eq:strm_fnct}
\end{equation}
This result can also be directly verified: acting with $(\partial_y^2 -k^2) $ on the right hand side of (\ref{eq:strm_fnct}), using that $\partial_y\omega_{k}^l\left(y-\frac{Y}k,t\right)=-k \partial_Y \omega_{k}^l\left(y-\frac{Y}k,t\right)$, integrating by parts twice, one gets $\omega_{k}^l\left(y,t\right)$ from the two boundary term containing $\partial_Y e^{-|Y|}$ taken at $Y=0$, while all other contributions either cancel or equal to zero.
The solution to (\ref{eq:w_fluct_min_app}) in Fourier space is then:
\begin{equation}
\omega_{k}^l(y,t)=\int_{-T}^t e^{ik U(y)(t'-t)}g_{kl}(y,t')e^{ily}dt'
\end{equation} 
where the steady state result is achieved once $T\to \infty$ . 
Using that $u=-\partial_y \psi$ and $v=\partial_x \psi$  we have

\begin{equation}
\begin{split}
\langle v_k^l(y) u_{-k}^{l'}(y)\rangle =\frac{i}{4k^3}\langle\int dY e^{-|Y|}\int_{-\infty}^{t}dt' e^{ik U(y-\frac{Y}k)(t-t')} g_k^l\left(y-\frac{Y}k,t'\right) \\\times\int dY' e^{-|Y'|}\int_{-\infty}^{t}dt'' \partial_y \left[e^{ik' U(y+\frac{Y'}{k})(t-t'')} g_{-k}^{l'}\left(y+\frac{Y'}k,t''\right)\right]\rangle
\end{split}
\end{equation} 
Changing $\partial_y\to+k\partial_{Y'}$ in the second line and integrating by parts (note that the boundary terms vanish at $Y'\to \pm\infty$ due to the exponential factors and cancel between $Y'>0$ and $Y'<0$ at $Y'=0$.)
\begin{equation}
\begin{split}
\langle v_k^l(y) u_{-k}^{l'}(y)\rangle =\frac{i}{4k^2}\int dY e^{-|Y|}\int dY' \left[\partial_{Y'}e^{-|Y'|}\right]\int_{-\infty}^{t}dt' \int_{-\infty}^{t}dt'' \langle g_k^l\left(y-\frac{Y}k,t'\right) g_{-k}^{l'}\left(y+\frac{Y'}k,t''\right)\rangle \\ \times e^{-ik U(y+\frac{Y'}{k})(t''-t)+ik U(y-\frac{Y}k)(t'-t)}. 
\end{split}
\end{equation} 
Averaging over the forcing and changing variables to $\tau=t'-t$ we get
\begin{equation}
\langle v_k^l(y) u_{-k}^{l'}(y)\rangle =\frac{i\eta_{k,ll'}e^{iy(l+l')} }{2k^2}\int_{-\infty}^{0} d\tau\int dY e^{-|Y|}\int dY' \left[\partial_{Y'}e^{-|Y'|}\right]   e^{-i\frac{l}k Y} e^{i\frac{l'}k Y'}e^{-ik U(y+\frac{Y'}{k})\tau}e^{ik U(y-\frac{Y}k)\tau}. 
\end{equation} 

The shear approximation $U \partial_{x} \omega  \gg -v U'' $ is valid for modes such that $U k \gg U'' k^{-1}$. This gives the condition $kd\gg1$ ($U'\sim U/d$), which for the forcing modes $k^2+l^2 \approx q_f^2$ is approximately equivalent to a restriction on the possible angles $-\Lambda \leq  \tan\theta \leq \Lambda$ with $\Lambda= q_f d$ and $l/k=\tan\theta$. We will assume the limit $\Lambda \to \infty$, such that almost all modes satisfy the former condition. We can then expand $U(y-\frac{Y}{k})\approx U(y)-\frac{Y}k U'(y)$, as long as $U'\neq 0$, since the velocity changes on a scale of the order $d\gg1/k$ and $Y,Y'\gg1 $ are exponentially suppressed in the integrals.
We therefore get

\begin{equation}
\langle v_k^l(y) u_{-k}^{l'}(y)\rangle =\frac{i\eta_{k,ll'}e^{iy(l+l')} }{2k^2}\int_{-\infty}^{0} d\tau \int dY e^{-|Y|}\int dY' \left[\partial_{Y'}e^{-|Y'|}\right]   e^{-i(\frac{l}k+U'\tau) Y} e^{-i(-\frac{l'}k+U'\tau) Y'}
\end{equation} 

which after performing the integrals over $Y$ and $Y'$ results in

\begin{equation}
\begin{split}
\langle v_k^l(y) u_{-k}^{l'}(y)\rangle =-\frac{2\eta_{k,ll'} }{k^2}e^{iy(l+l')}\int_{-\infty}^{0} d\tau \frac{-\frac{l'}{k}+U'\tau}{(1+(-\frac{l'}{k}+U'\tau)^2)(1+(\frac{l}{k}+U'\tau)^2))}\\ 
=2\eta_{k,ll'} e^{iy(l+l')}\int_{-\infty}^{0} d\tau \frac{k(l'-U'\tau k)}{(k^2+(l'-U'\tau k)^2)(k^2+(l+U'\tau k)^2))}
\end{split}
\end{equation} 

It is convenient to work with an expression which is more symmetric in $l,l'$. Also, physically, the result of excitation of the modes with $(k,l)$ and $(-k,l')$ by the forcing is the two contributions $\langle u_k^l(y) v_{-k}^{l'}(y)\rangle$ and $\langle u_{-k}^{l'}(y) v_{k}^{l}(y)\rangle$ to $\langle uv\rangle$. 
Using $\eta_{-k,l'l}=\eta_{k,ll'}$ we can write
\begin{equation}
\begin{split}
\frac{1}{2}\langle v_k^l(y) u_{-k}^{l'}(y)\rangle+\langle v_{-k}^{l'}(y) u_{k}^{l}(y)\rangle 
  =\eta_{k,ll'}e^{iy(l+l')}\int_{-\infty}^{0} d\tau \frac{k(l'-l-2U'\tau k)}{(k^2+(l'-U'\tau k)^2)(k^2+(l+U'\tau k)^2))}\\
 =-\frac{\eta_{k,ll'}e^{iy(l+l')}}{k^2}\int_{-\infty}^{0} d\tau \frac{\frac{l}{k}-\frac{l'}{k}+2U'\tau}{(1+(-\frac{l'}{k}+U'\tau)^2)(1+(\frac{l}{k}+U'\tau)^2))}
\end{split}
\end{equation}
Nicely, the integrand on the left hand side now simplifies to
\begin{equation}
\int_{-\infty}^{0} d\tau \frac{\frac{l}{k}-\frac{l'}{k}+2U'\tau}{(1+(-\frac{l'}{k}+U'\tau)^2)(1+(\frac{l}{k}+U'\tau)^2))}=\frac{1}{\frac{l}{k}+\frac{l'}{k}}\int_{-\infty}^{0} d\tau \left(\frac{1}{(1+(-\frac{l'}{k}+U'\tau)^2)}-\frac{1}{(1+(\frac{l}{k}+U'\tau)^2)}\right)
\end{equation}

Performing the integration we get
\begin{equation}
\frac{1}{2}\langle v_k^l(y) u_{-k}^{l'}(y)\rangle+\langle v_k^{-l'}(y) u_{-k}^{-l}(y)\rangle =\frac{\eta_{k,ll'}}{U'}e^{iy(l+l')}\frac{\arctan(l/k)+\arctan(l'/k)}{k(l+l')}. 
\end{equation} 
Finally, integrating over all the Fourier modes we obtain 
\begin{equation}
\begin{split}
\langle u(y)v(y)\rangle 
= \frac{1}{U'(y)} \int \frac{dl}{2\pi}\int \frac{dl'}{2\pi}\int \frac{dk}{2\pi}e^{iy(l+l')}\eta_{k,ll'}\frac{\arctan(l/k)+\arctan(l'/k)}{(l+l')k}
\label{uv_nh_general}
\end{split}
\end{equation}

Since the total energy injection $\bar{\epsilon}=1/L\int dy \epsilon(y)$ is non-zero, there will always be the contribution coming from the correlation between the modes  with $l'=-l$, and using $\frac{\arctan(l/k)+\arctan(l'/k)}{l/k+l'/k} \to \frac{1}{(1+(l/k)^2}$  we obtain the contribution
\begin{equation}
\langle u(y)v(y)\rangle = \frac{1}{U'(y)} \int \frac{dl}{2\pi}\int \frac{dk}{2\pi}\frac{\eta_{k,l,-l}}{k^2}=\frac{\bar{\epsilon}}{U'(y)}
\label{eq:uvA}
\end{equation}
for such modes, as previously.
Let us discuss the qualitative change for an inhomogeneous pumping so that $\epsilon(y)-\bar{\epsilon}=\delta\epsilon \neq 0$.
We change variables in (\ref{uv_nh_general}) to $l+l'=2s$, $l=s+j$, $l'=s-j$:
\begin{equation}
\begin{split}
\langle u(y)v(y)\rangle = 
\frac{1}{U'(y)} \int \frac{dj}{2\pi}\int \frac{2ds}{2\pi}\int \frac{dk}{2\pi}e^{i2ys}\eta_{k,s+j,s-j}\frac{\arctan((s+j)/k)+\arctan((s-j)/k)}{2sk}
\end{split}
\end{equation}

\subsection{The dominance of $k\approx 0$ modes - general considerations}
\label{sec:appB1}
Let us describe how the logarithmic contribution to the Reynolds stress arises . We will analyze an inhomogeneous contribution to the enstrophy covariance of the form $h(y+y')\chi((\bm{x-x'})/l_f)$. The function $\chi((\bm{x-x'})/l_f)$ controls the forcing correlation length, set to $l_f$, and $h(y)$ determines the variation in space. We assume that correlations decay fast at scales larger than $l_f$, such that, in Fourier space, $\chi((\bm{x-x'})/l_f)$ is approximately a delta function around $q_f$ (but the amplitude may vary with the direction).
Then, a Fourier harmonic of $h(y)$ with wavenumber $s$ contributes a mode with $l+l'= 2s$ to the enstrophy covariance. In addition, it shifts $\chi(\frac{\bm{x-x'}}{l_f})$ in Fourier space such that modes are roughly restricted to lie on the circle $(l-s)^2+k^2=q_f^2$. A convenient way to parameterize the excited modes is by $(k,l)=(k,s+j)$ such that $j^2+k^2\approx q_f$, i.e a circle with the origin at $(0,s)$. Then, a mode on the circle with a given $(k,j)$ is correlated with a mode with wavenumbers $(-k,-j)$, so that $l+l'=2s$.
The contribution from these modes to the momentum flux is given by:
\begin{equation}
\begin{split}
\langle v_k^{s+j}(y) u_{-k}^{s-j}(y)\rangle 
=2\eta_{k,s+j,s-j} e^{i2ys}\int_{-\infty}^{0} d\tau \frac{k(s-j-U'\tau k)}{(k^2+(s-j-U'\tau k)^2)(k^2+(s+j+U'\tau k)^2))}\\
=2 \eta_{k,s+j,s-j} e^{i2ys}\int_{-\infty}^{0} d\tau \frac{kl'(\tau)}{(k^2+(l'(\tau))^2)(k^2+(l(\tau))^2)}
\end{split}
\end{equation}

We see that the largest contribution would come from modes with $k\approx 0$ such that $l(\tau)=0, l'(\tau)\neq 0 $, i.e such that the initial excitation $(k,s+j)$ has been converted, due to the shear, to a mode with $(k,0)$. We see that for such modes $l'(\tau)=2s$ and that they exist only if $(j+s)/(U'k)>0$. They then contribute $\approx 4ks/(U'(k^2+4s^2)k^2)\eta_{k,s+j,s-j}$ which changes sign when $k$ changes sign. Note that we do not need to consider the contribution from $\langle u_{-k}^{s-j}(y) v_{k}^{s+j}(y)\rangle$ separately, since it is recovered by taking $k\to -k,j\to -j$ and recalling that $\eta_{k,l,l'}=\eta_{-k,l',l}$.
 
If $s<q_f$, since for $k\approx 0$ we have $j\approx q_f$ we get the condition $\frac{j}{U'k}>0$. However the contributions from $kU'>0$ ($j>0$) and $kU'<0$ $(j<0)$ cancel each other (because of the symmetry $\eta_{k,l,l'}=\eta_{-k,l',l}$). Therefore, eventually the modes with $k\approx 0$ play no significant role as their contribution cancels out. 

On the other hand, if $s>q_f$, the condition to have $l'(\tau)=0$ becomes $s/(U'k)>0$. Then, if $s>0$ then $U'k>0$ and $\pm j$ gives a contribution equal to  $ 4ks/(U'(k^2+4s^2)k^2)\eta_{k,s+j,s-j}\approx 1/(U'sk)\eta_{k,s+j,s-j}$. While if $s<0$ then $U'k<0$ and $\pm j$ gives a contribution equal to  $ 4ks/(U'(k^2+4s^2)k^2)\eta_{k,s+j,s-j}\approx 1/(U'sk)\eta_{k,s+j,s-j}$. 

In other words, for $k\approx 0$ and any $s\neq 0$, a contribution proportional to $1/k$ comes from $\tau$ such that $q(\tau)\approx k^2$---i.e $l(\tau)\approx 0$ and $l'(\tau)\approx 2s\neq 0$. It therefore has the same sign as $(ks)\eta_{k,ll'}$. 

The difference between $s<q_f$ and $s>q_f$ has to do with the type of modes which can have $l(\tau)=0$.
For $s<q_f$, in wavenumber space, the circle of modes initially excited by the forcing contains the origin. There are therefore always two modes with $l(\tau)=0$. For $U'>0$ ($U'<0$), they originate either from the first or the third quadrant: $j/k>0$ ($j/k<0$). Then, for a pair of correlated modes, if one of them has some excitation time $\tau$ such that $l(\tau)=0$, there will be another excitation time $\tau'$, such that the second mode has $l(\tau')=0$ (giving a $1/k$ contribution to $\langle v_{-k}^{l'}(y) u_{k}^{l'}(y)\rangle$). These two contributions to $\langle uv\rangle$ are equal with opposite signs, since two correlated modes come with opposite signed $k$, and thus cancel out. 
 
On the other hand, if $s>q_f$ the origin lies outside the circle of excited modes. Modes with $l(\tau)=0$ develop once the circle is elongated enough by shear (forming an ellipse) such that it intersect the $k$ axis. Modes with $l(\tau)=0$ originate from half the plane, having $k>0$ ($k<0$) for $s>0$ ($s<0$) and $U'>0$. Thus, for each correlated pair of modes, only on of them (if at all) can develop $l(\tau)=0$ for some excitation time. Thus, generically, cancellations do not occur. 

We can rewrite it in terms of the energy  
\begin{equation}
\eta_{k,s+j,s-j}=\epsilon_{k,s+j,s-j}\frac{((s+j)^2+k^2)((s-j)^2+k^2)}{k^2+j^2-s^2}\approx -\frac{1}{s^2}\epsilon_{0,s,s}
\end{equation}
so that $\eta_{k,s+j,s-j}\approx \eta_{0,s,s}=-\frac{1}{s^2}\epsilon_{0,s,s}$
for $s\gg q_f, k\approx 0, j\approx q_f$. Note that even for $q_f \lesssim s$ the sign of $\epsilon_{0,s,s}$ and $\eta_{0,s,s}$ are opposite. For $s\gg q_f$ this finally gives the contribution
\begin{equation}
\langle v_k^{s+j}(y) u_{-k}^{s-j}(y)\rangle+\langle v_k^{-s+j}(y) u_{-k}^{-s-j}(y)\rangle=-\left(e^{i2ys}\epsilon_{0,s,s}+e^{-i2ys}\epsilon_{0,-s,-s}\right) \frac{s}{k U'}
\end{equation}
for $k\approx 0$.

This is the dominant contribution, which gives the $\log \Lambda$ factor. The minus sign probably comes from the fact that the modes that can turn to zero have the orthogonal direction to the one the shear tends to align modes with.

\subsection{Model for enstrophy gradient much larger than the correlation wavenumber}
\label{sec:appB2}
As a model for an inhomogeneous pumping in the $y$ direction we take the vorticity covariance to be of the form $\langle g(\bm{x})g(\bm{x'})\rangle= 2\bar{\eta}\cos sy \cos sy' J_0(q_f r)2\delta(t+t')=\bar{\eta}(\cos s(y+y')+ \cos s(y-y')) J_0(q_f r)2\delta(t+t')$ where $r^2=(x-x')^2+(y-y')^2$ and $J_0$ is the Bessel function of order zero and $s>0$. For $s=0$ the forcing is both homogeneous and isotropic with correlation length equal to $q_f$.  
We take $s \gg q_f$, so that the characteristic gradient of $\eta(y)=\bar{\eta}(1+\cos 2sy)$ is much larger than $q_f$, the characteristic correlation length of the forcing \footnote{We can also consider a more general form, non-isotropic at $s=0$, \unexpanded{$\langle g(\bm{x})g(\bm{x'})\rangle= 2\bar{\eta}\cos sy \cos sy' \sum a_{2n} J_{2n}(q_f r)\cos( 2n \phi)+b_{2n} J_{2n}(q_f r)\sin( 2n \phi) 2\delta(t+t')$} where $\phi$ denotes the angle between $\bm{x-x'}$ and the $x$-axis.}.

We then have
\begin{equation}
\begin{split}
\eta_{k,ll'}=\frac{1}2 \left[ 2\pi \delta(l+l')+2\pi \delta(l+l'-2s)\right]\frac{\bar{\eta}}{q_f}2\pi\delta\left(\sqrt{k^2+(l-s)^2}-q_f\right)+\\
+\frac{1}2 \left[ 2\pi \delta(l+l')+2\pi \delta(l+l'+2s)\right]\frac{\bar{\eta}}{q_f}2\pi\delta\left(\sqrt{k^2+(l+s)^2}-q_f\right)
\end{split}
\end{equation}
using that the Fourier transform of the Bessel function is ${\cal F} [J_0(q_f r)] =\delta(l+l')\frac{2\pi}{q_f} \delta\left(\sqrt{k^2+l^2}-q_f\right)$ and that the factor $e^{isy}e^{\pm isy'}$ acts as a shift in Fourier space: $l\to l-s$, $l'\to l' \mp s$.

We begin with the form
\begin{equation}
\begin{split}
\langle u(y)v(y)\rangle = \frac{1}{U'(y)} \int \frac{dl}{2\pi}\int \frac{dl'}{2\pi}\int \frac{dk}{2\pi}e^{iy(l+l')}\eta_{k,ll'}\frac{\arctan(l/k)+\arctan(l'/k)}{k(l+l')}
\end{split}
\end{equation}

For the part containing $\delta(l+l')$ we have
\begin{equation}
\begin{split}
 \frac{1}{U'(y)} \int \frac{dl}{2\pi}\int \frac{dl'}{2\pi}\int \frac{dk}{2\pi}\frac{\bar{\eta} }{l^2+k^2}\frac{1}2\left(\frac{2\pi}{q_f}\delta\left(\sqrt{k^2+(l-s)^2}-q_f\right)+\frac{2\pi}{q_f}\delta\left(\sqrt{k^2+(l+s)^2}-q_f\right)\right)= \frac{\bar{\eta}}{U'(y)}\frac{1}{s^2-q_f^2}.
 \label{eq:delta0}
\end{split}
\end{equation}
For the factor $\delta(l+l'-2s)$ we have
\begin{equation}
\begin{split}
\frac{1}2\frac{\bar{\eta}}{q_f}\frac{e^{i2ys}}{U'(y)} \int \frac{dl}{2\pi}\int dk\delta\left(\sqrt{k^2+(l-s)^2}-q_f\right)\frac{\arctan(l/k)+\arctan((2s-l)/k)}{2sk}=\\
=\frac{1}2\frac{\bar{\eta}}{sq_f}\frac{e^{i2ys}}{U'(y)} \int \frac{d\tilde{l}}{2\pi}\int dk\delta\left(\sqrt{k^2+\tilde{l}^2}-q_f\right)\frac{\arctan((\tilde{l}+s)/k)+\arctan((s-\tilde{l})/k)}{2k}=\\
=\frac{1}2\frac{\bar{\eta}}{sq_f}\frac{e^{i2ys}}{U'(y)} \int q dq \delta\left(q-q_f\right) \int\frac{d\theta}{2\pi}\frac{\arctan((q\sin\theta+s)/q\cos\theta)+\arctan((s-q\sin\theta)/q\cos\theta)}{2q\cos\theta}=\\
=\frac{1}2\frac{\bar{\eta}}{sq_f}\frac{e^{i2ys}}{U'(y)} \int\frac{d\theta}{2\pi}\frac{\arctan((q_f\sin\theta+s)/q_f\cos\theta)+\arctan((s-q_f\sin\theta)/q_f\cos\theta)}{2\cos\theta} \approx \\
\approx \frac{1}2\frac{\bar{\eta}}{sq_f}\frac{e^{i2ys}}{U'(y)} \int\frac{d\theta}{2\pi}\frac{\arctan(\frac{s}{q_f\cos\theta})}{\cos\theta}
\end{split}
\end{equation}
 where in the second line we changed variables to $\tilde{l}=l-s$ and in the line before last the fact that $s\gg q_f$. In the limit $s\gg q_f$ we have
\begin{equation}
\begin{split}
\int \frac{d\theta}{2\pi}\frac{\arctan(\frac{s}{q\cos\theta})}{\cos\theta}&=\text{sign(s)}\frac{\pi}2 \left(\int_{-\pi/2+\Lambda^{-1}}^{\pi/2-\Lambda^{-1}} \frac{1}{\cos\theta}\frac{d\theta}{2\pi}-\int_{\pi/2+\Lambda^{-1}}^{3\pi/2-\Lambda^{-1}} \frac{1 }{\cos\theta}\frac{d\theta}{2\pi}\right)
=\text{sign(s)}\pi\int_{-\pi/2+\Lambda^{-1}}^{\pi/2-\Lambda^{-1}} \frac{1}{\cos\theta}\frac{d\theta}{2\pi}=\\
&=\text{sign(s)}\int_{0}^{\pi/2-\Lambda^{-1}}  \frac{d\theta}{\cos\theta}
=\text{sign(s)}\left.\log \left(\frac{\sin \left(\frac{\theta }{2}\right)+\cos \left(\frac{\theta }{2}\right)}{\cos \left(\frac{\theta }{2}\right)-\sin \left(\frac{\theta }{2}\right)}\right)\right|_{0}^{\pi/2-\Lambda^{-1}}\xrightarrow[\Lambda\to \infty]{}\text{sign(s)}\log\Lambda
\end{split}
\end{equation} 
so that we get the contribution
\begin{equation}
\frac{1}2\frac{\bar{\eta}}{q_f}\frac{e^{i2ys}}{U'(y)}\frac{\text{sign(s)}}s\log\Lambda
\end{equation}
from  $\delta(l+l'-2s)$. The contribution from $\delta(l+l'+2s)$ is obtained by changing $s\to-s$ in the above result giving
\begin{equation}
\langle uv\rangle \approx\frac{1}2\frac{\bar{\eta}}{sq_f}\frac{e^{i2ys}}{U'(y)}\log\Lambda+\frac{1}2\frac{\bar{\eta}}{sq_f}\frac{e^{i2ys}}{U'(y)}\log\Lambda+\frac{\bar{\eta}}{U'(y)}\frac{1}{s^2}=\frac{\bar{\eta}}{U'(y)}\frac{1}{s^2}+\cos 2sy \frac{\bar{\eta}}{U'(y)}\frac{\log\Lambda}{sq_f}
\end{equation}
keeping only the leading order in $q_f/s$.

It is most natural to express $\langle uv\rangle$ in terms of $\delta\epsilon(y)=\epsilon(y)-\bar{\epsilon}$ and $\bar{\epsilon}=\int dy \epsilon(y)$ is the zeroth mode of $\epsilon(y)$. Recall that $\epsilon_{k,ll'}=(k^2-ll')/\left[(k^2+l^2)(k^2+l'^2)\right]\eta_{k,ll'}$. The terms in $\eta_{k,ll'}$ containing $\delta(l+l')$ give $(k^2-ll')/\left[(k^2+l^2)(k^2+l'^2)\right]=1/{k^2+l^2}$ so that $\bar{\epsilon}=\frac{\bar{\eta}}{s^2-q_f^2}$, as we may expect due to the result (\ref{eq:delta0}) and that $\langle uv\rangle=\epsilon/U'$ for a homogeneous pumping. Let us compute the contribution from $\delta(l+l'-2s)$ to $\epsilon(y)$:

\begin{equation}
\begin{split}
\frac{1}2\frac{\bar{\eta}}{q_f}e^{i2ys}\int \frac{dl}{2\pi}\int dk\delta\left(\sqrt{k^2+(l-s)^2}-q_f\right)\frac{k^2-l(2s-l)}{(k^2+l^2)(k^2+(2s-l)^2)}=\\
=\frac{1}2\frac{\bar{\eta}}{q_f}e^{i2ys} \int \frac{d\tilde{l}}{2\pi}\int dk\delta\left(\sqrt{k^2+\tilde{l}^2}-q_f\right)\frac{k^2-(\tilde{l}+s)(s-\tilde{l})}{(k^2+(\tilde{l}+s)^2)(k^2+(s-\tilde{l})^2)}=\\
=\frac{1}2\frac{\bar{\eta}}{q_f}e^{i2ys}\int q dq \delta\left(q-q_f\right) \int\frac{d\theta}{2\pi}\frac{q^2-s^2}{(q^2+2sq\sin\theta+s^2)(q^2-2sq\sin\theta+s^2)}=-\frac{1}2e^{i2ys}\frac{\bar{\eta}}{q_f^2+s^2}
\end{split}
\end{equation}
 so that we have for the local energy injection rate
 \begin{equation}
 \epsilon(y)=\frac{\bar{\eta}}{s^2-q_f^2}- \cos 2sy\frac{\bar{\eta}}{s^2+q_f^2}\approx \frac{\bar{\eta}}{s^2}(1-\cos 2sy).
\end{equation}  

We then have
\begin{equation}
\langle uv\rangle=\frac{\bar{\epsilon}}{U'}\left(1+\cos 2sy\frac{s}{q_f}\log\Lambda\right) =\frac{\bar{\epsilon}}{U'}-\frac{\delta\epsilon(y)}{U'}\frac{s}{q_f}\log\Lambda
\end{equation}

\section{Direct check of $\langle vp\rangle =0 $ for a homogeneous pumping}
\label{sec:vp}
To leading order the fluctuating part of the pressure satisfies the equation
\begin{equation}
-(\partial_y^2-k^2)p_k^l=2 U' i k v_k^l
\end{equation} 
and 
\begin{equation}
\langle v_k^l p_{-k}^{l'}\rangle =\frac{i}{k}\int_{-\infty}^{\infty}dY e^{-|Y|}U'(y+Y/k)\left\langle v_{k}^l(y)v_{-k}^{l'}\left(y+\frac{Y}k,t\right)\right\rangle
\end{equation}
\begin{equation}
\begin{split}
\langle v_k^l(y) v_{-k}^{l'}(y')\rangle = -\frac{\eta_{k,ll'}e^{i(yl+l'y')} }{2k^2}\int_{-\infty}^{0} d\tau \\ \times\int dX e^{-|X|}\int dY' e^{-|Y'|} e^{i k(U(y-\frac{X}{k})-U(y'+\frac{Y'}{k}))\tau}  e^{-i\frac{l}k X} e^{i\frac{l'}k Y'}
\end{split}
\end{equation} 

\begin{equation}
\langle v_k^l p_{-k}^{l'}\rangle =-\frac{i\eta_{k,ll'}}{2k^3}e^{iy(l+l))}\int_{-\infty}^{\infty}dY e^{-|Y|}U'(y-Y/k)e^{i\frac{l'}{k}Y}\int_{-\infty}^{0} d\tau \int dX e^{-|X|}\int dY' e^{-|Y'|} e^{ik (U(y-\frac{X}{k})-U(y+\frac{Y}{k}+\frac{Y'}{k}))\tau}  e^{-i\frac{l}k X} e^{i\frac{l'}k Y'}
\end{equation}
which to leading order reads
\begin{equation}
\langle v_k^l p_{-k}^{l'}\rangle =-\frac{iU'(y)\eta_{k,ll'}}{2k^3}e^{iy(l+l'))}\int_{-\infty}^{0} d\tau\int_{-\infty}^{\infty}dY e^{-|Y|} \int dX e^{-|X|}\int dY' e^{-|Y'|}  e^{i -U'(y)(X+Y+Y')\tau}  e^{-i\frac{l}k X} e^{i\frac{l'}k (Y'+Y)}
\end{equation}
Now for a homogeneous pumping:
\begin{equation}
\begin{split}
\langle v_k^l p_{-k}^{-l}\rangle =-\frac{iU'(y)\eta_{k,l}}{2k^3}\int_{-\infty}^{0} d\tau\int_{-\infty}^{\infty}dY e^{-|Y|} \int dX e^{-|X|}\int dY' e^{-|Y'|}  e^{-i U'(y)(X+Y+Y')\tau}  e^{-i\frac{l}k( X+Y'+Y)} \\=
-\frac{iU'(y)\eta_{k,l}}{2k^3}\int_{-\infty}^{0} d\tau\int_{-\infty}^{\infty}dY e^{-|Y|} \int dX e^{-|X|}\int dY' e^{-|Y'|} \cos\left( (U'(y)\tau+\frac{l}k)(X+Y+Y')\right) 
\end{split}
\end{equation}
where we have used that the anti symmetric part of the integral with respect to $(X,Y,Y')\to (-X,-Y,-Y')$ integrates to zero. 
Since $\eta_k,l$ is real, $\langle v_k^l p_{-k}^{-l}\rangle$ is purely imaginary and  $\langle v(y) p(y)\rangle=0$.
\end{widetext}

\acknowledgments
I am grateful to G. Falkovich and anonymous referee 2 for discussions that improved this work. I am also indebted to N. Constantinou, T. Drivas, D. Leocanet and G. Falkovich for a careful reading of the manuscript. I wholeheartedly thank Y. Hammer.  This work was funded by Princeton University and the Israel National Postdoctoral Award Program for Advancing Women in Science.

%

\end{document}